\documentclass[
    aps,
    prx,
    reprint,
    amsmath,
    amssymb,
    superscriptaddress,
    longbibliography
]{revtex4-2}

\usepackage{graphicx}
\usepackage{dcolumn}
\usepackage{bm}
\usepackage{hyperref}
\usepackage{xcolor}
\usepackage[dvipsnames]{xcolor}
\usepackage{booktabs}
\usepackage{float}

\usepackage{mathrsfs}
\usepackage{braket}
\usepackage[utf8]{inputenc}
\usepackage{textcomp}
\usepackage{textgreek}
\usepackage[normalem]{ulem}

\hypersetup{
    colorlinks = true,
    linkcolor  = blue,
    citecolor  = blue,
    urlcolor   = blue
}

\definecolor{ddgreen}{RGB}{0,120,70}

\usepackage[color=orange!60,textsize=footnotesize]{todonotes}
\begin{document}

\title{Approaching the Inverse Neutron Scattering Problem with Neural Networks}

\author{Jingyi Luo}
\email[Corresponding author: ]{ljy.ds8@gmail.com}
\affiliation{Department of Physics and Astronomy, University of Tennessee, Knoxville, Tennessee 37996, USA}

\author{Fletcher Williams}
\affiliation{Department of Physics and Astronomy, University of Tennessee, Knoxville, Tennessee 37996, USA}

\author{Ang-Kun Wu}
\affiliation{Department of Physics and Astronomy, University of Tennessee, Knoxville, Tennessee 37996, USA}

\author{David Dahlbom}
\affiliation{Neutron Scattering Division, Oak Ridge National Laboratory, Oak Ridge, Tennessee 37831, USA}

\author{Cristian D. Batista}
\affiliation{Department of Physics and Astronomy, University of Tennessee, Knoxville, Tennessee 37996, USA}
\affiliation{Neutron Scattering Division, Oak Ridge National Laboratory, Oak Ridge, Tennessee 37831, USA}

\author{Hao Zhang}
\email[Corresponding author: ]{hzhang68@utk.edu}
\affiliation{Department of Physics and Astronomy, University of Tennessee, Knoxville, Tennessee 37996, USA}

\date{\today}

\begin{abstract}
Determining the microscopic Hamiltonian of a quantum magnet from inelastic neutron scattering measurements remains a central challenge in condensed matter physics. While the dynamical spin structure factor contains, in principle,  sufficient information about the underlying interactions, extracting Hamiltonian parameters from experimental spectra constitutes a highly nontrivial inverse problem due to the large parameter space and the presence of experimental noise. Here we demonstrate that  neural networks can efficiently solve this inverse problem for a broad class of quantum magnets. As a benchmark, we consider an eight-parameter family of honeycomb lattice Heisenberg spin models  in the fully polarized phase, where the dynamical spin structure factor can be computed exactly within linear spin-wave theory. Using a large synthetic dataset of neutron scattering spectra, we train neural networks to infer the underlying Hamiltonian parameters directly from the dynamical response. We compare the performance of three different architectures---fully connected neural networks (FCNNs), one-dimensional convolutional neural networks (CNN1Ds), and two-dimensional convolutional neural networks (CNN2Ds)---and find that all achieve high predictive accuracy. Remarkably, the trained models remain robust in the presence of substantial experimental uncertainty, reliably recovering the Hamiltonian parameters even when the input spectra are contaminated by random noise with amplitudes reaching $10\%$ of the signal intensity. Our results establish machine learning as a powerful framework for quantitative Hamiltonian reconstruction from neutron scattering data and provide a practical route toward automated characterization of quantum magnetic materials.
\end{abstract}

\maketitle

\section{Introduction}

Determining the microscopic Hamiltonian of a quantum material from experimental observations is a central objective of condensed matter physics. In magnetic systems, the Hamiltonian encodes the exchange interactions, anisotropies, and competing energy scales that govern collective spin behavior, providing the foundation for understanding thermodynamic properties, excitation spectra, phase transitions, and responses to external perturbations. Reconstructing these interactions is therefore essential for connecting experimental observations to the microscopic mechanisms underlying material behavior. Inelastic neutron scattering (INS) offers a powerful probe through the dynamical spin structure factor $S(\mathbf q,\omega)$. Modern spectrometers measure magnetic excitations over extended regions of momentum and energy space, producing rich datasets that encode the underlying interactions. Turning this spectral information into quantitative microscopic Hamiltonians, however, remains a challenging inverse problem.

The conventional approach relies on forward modeling: the parameters of a candidate Hamiltonian are refined iteratively by comparing its calculated dynamical response with experimental measurements. Repeated forward calculations can be computationally demanding, while parameter correlations and multiple local minima complicate the optimization and assessment of the uniqueness of a solution.

Machine learning offers complementary approaches to this problem. Autoencoder-based latent representations, global optimization, and learned forward surrogates have been used to determine interaction parameters from diffuse and inelastic neutron-scattering data in $\mathrm{Dy_2Ti_2O_7}$ and $\alpha$-$\mathrm{RuCl_3}$~\cite{Samarakoon2020,Samarakoon2022RuCl3,Samarakoon2022Pressure}. Neural networks have also been used to distinguish competing exchange models in half-doped manganites~\cite{Butler2021}, translate between simulated and experimental INS spectra~\cite{Anker2023}, and construct a differentiable forward model for refining exchange parameters in $\mathrm{La_2NiO_4}$~\cite{Chitturi2023}. Active-learning strategies combining adaptive noise reduction, sparse data acquisition, and magnetic parameter extraction have been demonstrated for $\mathrm{CrSBr}$~\cite{Abuawwad2025}. Related inversion strategies have been used to extract interatomic force constants from powder and single-crystal phonon spectra~\cite{SuLi2024,Sable2026}, while networks trained on spectra generated by Landau--Lifshitz--Gilbert dynamics have been used to infer Hamiltonian parameters from terahertz spectroscopy~\cite{Mootz2026terahertz}.

Here we investigate a direct-inversion approach in which a neural network learns the mapping from $S(\mathbf q,\omega)$ to the parameters of a prescribed family of spin Hamiltonians. We generate synthetic spectra using linear spin-wave theory and use them to train networks for Hamiltonian reconstruction. Once trained, a network can infer parameters from previously unseen spectra without repeating the forward calculation or performing a separate optimization for each dataset.

Beyond accelerating inference, this approach provides a practical way to examine which spectral observables constrain the Hamiltonian. Theoretical results establish related uniqueness connections between static structure factors, the coupling constants of bilinear spin Hamiltonians, and ground-state wave functions~\cite{Murta2022,Quintanilla2022}. Comparing reconstruction from different inputs can expose ambiguities and identify the additional information needed to distinguish competing parameter sets, helping guide experimental design~\cite{TeixeiraParente2023}. In particular, excitation energies and spectral weights need not provide equivalent information about the microscopic interactions.

Reliable reconstruction also depends on the quality and coverage of the training data~\cite{mohammed2025effects,gong2023survey}. We therefore establish this approach in a controlled proof-of-principle setting: an eight-parameter family of honeycomb Heisenberg Hamiltonians in the fully polarized phase. In this regime, the one-magnon sector is described exactly by linear
spin-wave theory, yielding the exact one-magnon energies and
zero-temperature transverse dynamical spin structure factor. The
forward problem can therefore be solved both exactly and efficiently,
allowing us to isolate the challenges of inverse reconstruction from
uncertainties in the calculated spectra while retaining a nontrivial
eight-parameter model family.

Using the resulting synthetic database, we compare fully connected neural networks (FCNNs)~\cite{sazli2006brief,fine1999feedforward,bebis1994feed}, one-dimensional convolutional neural networks (CNN1Ds)~\cite{malek2018one,qazi2022one,khan2023detection}, and two-dimensional convolutional neural networks (CNN2Ds)~\cite{li2021survey,o2015introduction,krichen2023convolutional}. We examine their reconstruction accuracy, their dependence on the spectral information supplied, and their robustness to modeled experimental effects, including finite energy resolution and counting noise.

Beyond establishing a proof of principle, this work provides a framework for developing broadly accessible tools for Hamiltonian reconstruction from neutron-scattering data. Extending its reach will require forward solvers that are both accurate enough to describe the relevant dynamics and efficient enough to generate representative training datasets.

The remainder of this paper is organized as follows.
Section~\ref{sec:inverse} formulates the inverse neutron-scattering problem
and presents the exact forward solver for the model family considered here.
Section~\ref{sec:NN} describes the construction of the synthetic training
database and the neural-network architectures used to reconstruct
Hamiltonian parameters.
Section~\ref{sec:results} benchmarks reconstruction accuracy and robustness
and examines the complementary information provided by magnon dispersions
and spectral weights.
Section~\ref{sec:closing_loop} connects the idealized spectra to experimental
measurements by incorporating finite instrumental resolution and counting
noise and evaluating their effects on reconstruction.
Finally, Secs.~\ref{sec:discussion} and~\ref{sec:conclusion} discuss the
implications and limitations of the approach and summarize our main findings.

\section{Inverse scattering problem}
\label{sec:inverse}

The microscopic description of a family of magnetic materials can often be
restricted to a physically motivated class of Hamiltonians,
$
    H(\boldsymbol{\theta}),
$
where the parameter vector $\boldsymbol{\theta}$ contains the exchange
couplings and other microscopic interactions expected to be relevant for
that family of materials. For a given set of parameters
$\boldsymbol{\theta}$, a theoretical forward solver, denoted as $\mathcal{F}$, 
provides the corresponding zero-temperature
dynamical spin structure factor,
\[
    \boldsymbol{\theta}
    \xrightarrow{\;\mathcal F\;}
    S(\mathbf q,\omega),
\]
which determines the magnetic contribution to the neutron-scattering
response. This defines the forward scattering problem.

The inverse problem addressed in this work is the opposite one: given a
measured neutron-scattering response, determine the parameters
$\boldsymbol{\theta}$ of the Hamiltonian within the prescribed model
class. In contrast to the forward problem, for which powerful analytical
and numerical methods are available, this inverse problem is generally
much more difficult. In particular, different regions of parameter space
may produce very similar spectra, experimental data contain noise and
have finite resolution, and conventional parameter estimation requires
repeated evaluations of the forward solver while exploring a
multidimensional parameter space.

Our approach is to perform this computationally demanding exploration
only once. We sample the parameter space of the chosen Hamiltonian family
and use a forward solver to generate the corresponding neutron-scattering
responses. The resulting pairs
$
    \left\{\boldsymbol{\theta},
    S_{\boldsymbol{\theta}}(\mathbf q,\omega)\right\}
$
form a synthetic database that is used to train a neural network to learn
the inverse relation between the neutron response and the microscopic
parameters. Once trained, the network can be deployed on neutron-scattering
data from materials expected to belong to the same model family, providing
a direct estimate of their Hamiltonian parameters without repeatedly
solving the forward problem.

\begin{figure}
    \centering
    \includegraphics[width=0.48\textwidth]{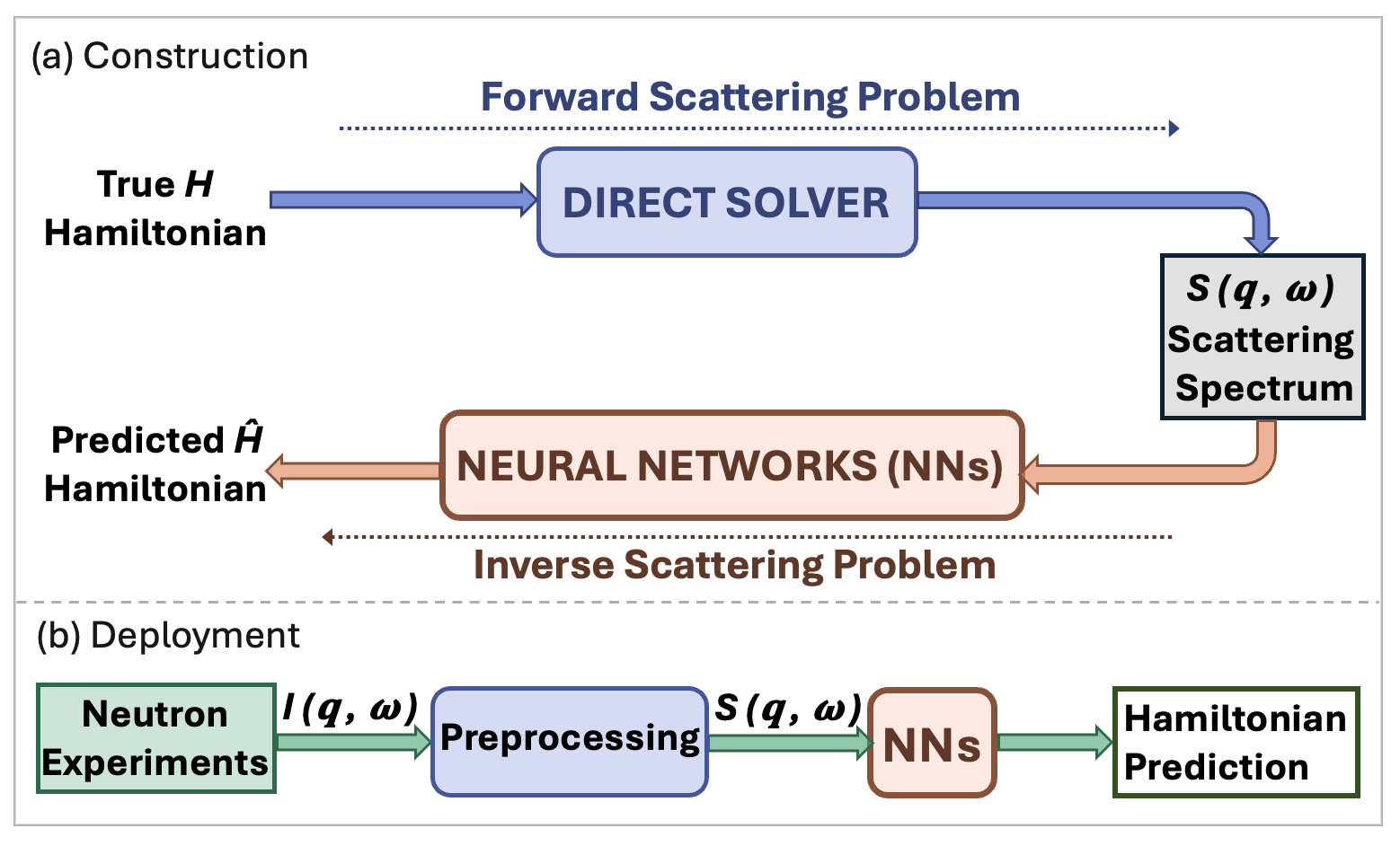}
    \caption{Schematic workflow for the construction and deployment of
    the inverse-scattering model. During construction, a forward solver
    generates neutron-scattering responses for a family of microscopic
    Hamiltonians, which are used to train the neural network. During
    deployment, the trained network infers the Hamiltonian parameters
    directly from neutron-scattering data.}
    \label{fig:flow_chart}
\end{figure}

This construction is summarized schematically in
Fig.~\ref{fig:flow_chart}. During the construction stage, the forward
solver encodes the microscopic physics of the chosen Hamiltonian family
into the training database, while the neural network learns the
corresponding inverse map. During deployment, the trained network takes
the neutron-scattering response as input and directly predicts the
Hamiltonian parameters.

The central requirement of this strategy is therefore the availability
of a forward solver that is sufficiently accurate and efficient to
generate a large training database over the relevant Hamiltonian
parameter space. To establish the method in a controlled setting, we
consider below a family of Hamiltonians for which the relevant
neutron-scattering response can be calculated exactly and at very low
computational cost.

\subsection{Forward scattering problem}
\label{subsec:forward}

\begin{figure}
    \centering
    \includegraphics[width=.38\textwidth]{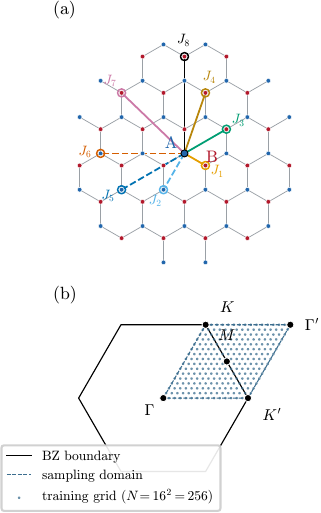}
    \caption{
    (a) Honeycomb lattice showing sublattices $A$ (blue) and $B$ (red),
    with nearest-neighbor bonds in gray. Colored lines mark one
    representative of each isotropic Heisenberg exchange coupling
    $J_1$--$J_8$ from a reference $A$ site; solid (dashed) lines connect
    opposite (same) sublattices.
    (b) Uniform $16\times16$ momentum grid (256 points) used to generate
    the training spectra. The dashed parallelogram, bounded by $\Gamma$,
    $K$, $\Gamma'$, and $K'$, marks the sampling domain; its area is one
    third that of the first Brillouin zone (solid boundary).
    }
    \label{fig:honeycomb}
\end{figure}

To establish the inverse-learning strategy in a controlled setting, we consider an eight-parameter family of isotropic Heisenberg models on the honeycomb lattice in the fully polarized phase. 
For this model family, the parameter vector introduced above is
\[
    \boldsymbol{\theta}=\{J_1,\ldots,J_8\},
\]
where $J_1,\ldots,J_8$ are the exchange couplings for the first through
eighth neighbor shells. 
Figure~\ref{fig:honeycomb}(a) shows the two honeycomb sublattices and one representative bond for each coupling.

The Hamiltonian for this model family is
\begin{equation}
    H=
    \frac12\sum_{{\bf R},{\bf r}}
    \sum_{\alpha,\beta=A,B}
    J_{\bf r}^{\alpha\beta}\,
    {\bf S}_{\bf R}^{\alpha}\cdot
    {\bf S}_{{\bf R}+{\bf r}}^{\beta}
    -h\sum_{{\bf R},\alpha}S_{\bf R}^{z,\alpha}.
    \label{eq:H}
\end{equation}
Here ${\bf R}$ labels a primitive cell, $\alpha$ and $\beta$ label the two honeycomb sublattices, and $J_{\bf r}^{\alpha\beta}$ is the isotropic exchange coupling between the indicated sites. 
Each $J_i$ is assigned to all bonds in its neighbor shell, as illustrated in Fig.~\ref{fig:honeycomb}(a). 
The parameter $h$ is the Zeeman energy associated with the applied field. 
We restrict attention to $h>h_{\rm sat}$, where the ground state is fully polarized. 
Acting on this state, a transverse spin-lowering operator creates a single magnon. 
The quadratic Holstein--Primakoff Hamiltonian is exact within the one-magnon sector, so it gives both the one-magnon dispersions and the zero-temperature transverse dynamical structure factor $S^{+-}({\bf q},\omega)$ exactly. 
Its low computational cost also permits the generation of a large training database across the eight-dimensional parameter space.

After a Holstein--Primakoff expansion about the fully polarized state (see Appendix~\ref{section:energies}), the spin-wave Hamiltonian takes the form
\begin{equation}
    H_{\rm sw}
    =\sum_{\bf q}
    {\bf b}_{\bf q}^{\dagger}M({\bf q}){\bf b}_{\bf q},
    \label{eq:Hsw}
\end{equation}
where
\begin{equation}
    {\bf b}_{\bf q}
    =
    \begin{pmatrix}
    b_{{\bf q},A}\\
    b_{{\bf q},B}
    \end{pmatrix}
    \label{eq:bvec}
\end{equation}
is the vector of Fourier-transformed Holstein--Primakoff bosons, and
\begin{equation}
    M({\bf q})=hI_2+S[J({\bf q})-D].
    \label{eq:Mq}
\end{equation}
Here $I_2$ is the $2\times2$ identity matrix, $J({\bf q})$ is the exchange matrix, and $D$ contains its exchange row sums. 
Their explicit forms, including the bond-sum convention used by the forward solver, are given in Appendix~\ref{section:energies}.

With two sites in the primitive unit cell, we obtain two magnon bands:
\begin{equation}
    \omega_\pm({\bf q})=h+S\mu_\pm({\bf q}),
    \label{eq:magnon_dispersion}
\end{equation}
where
\begin{equation}
    \mu_\pm({\bf q})=
    2\operatorname{Re}J^{AA}({\bf q})
    -\left[2J^{AA}({\bf 0})+J^{AB}({\bf 0})\right] \pm|J^{AB}({\bf q})|.
\end{equation}
This expression uses the equivalence of the $A$ and $B$ sublattices. 
The fully polarized state is stable as long as both magnon energies are positive at every momentum. 
The instability first occurs at the wave vector and band that minimize $\mu_n({\bf q})$. 
Hence the saturation field is
\begin{equation}
    h_{\rm sat}=-S\min_{{\bf q},n}\mu_n({\bf q}).
\end{equation}

In the fully polarized phase, only one-magnon states contribute to the inelastic transverse dynamical structure factor. 
With the per-spin normalization derived in Appendix~\ref{section:DSSF}, the result is
\begin{equation}
    S^{+-}({\bf q},\omega)
    =\sum_{n=\pm}I_n({\bf q})
    \delta\!\left[\omega-\omega_n({\bf q})\right],
    \label{eq:Sqw}
\end{equation}
with spectral weights
\begin{equation}
    I_\pm({\bf q})
    =S\left[
    1\pm\cos\!\left(
    2\pi{\bf q}\cdot{\bf d}+\phi_{\bf q}
    \right)\right].
    \label{eq:intensity}
\end{equation}
Here ${\bf d}={\bf d}_B-{\bf d}_A$ is the displacement between the two sites within the primitive cell, and
$\phi_{\bf q}=\arg J^{AB}({\bf q})$. 
The interference term combines the basis-position phase with the relative phase of the magnon eigenvector. 
Thus the spectral weights contain information about the exchange couplings beyond that contained in the magnon energies.

The magnon energies and spectral weights define the exact forward map
\begin{equation}
    {\cal F}:\{J_1,\ldots,J_8\}
    \longrightarrow S^{+-}({\bf q},\omega)
    \label{eq:forward_map}
\end{equation}
from the eight-dimensional Hamiltonian parameter space to the zero-temperature transverse dynamical structure factor. 
The inverse problem is to determine to what extent the couplings $\{J_i\}$ can be reconstructed from this response.

\subsection{Inverse problem}
\label{subsec:inverse}

Conventional Hamiltonian reconstruction proceeds by solving the inverse problem through iterative optimization. 
Starting from a trial set of microscopic parameters, the forward problem is solved and the resulting dynamical response is compared with the experimental data. 
The parameters are then adjusted and the procedure is repeated until satisfactory agreement is obtained. This approach can become computationally demanding when the forward solver is expensive and the Hamiltonian contains many unknown parameters. 
Moreover, strong parameter correlations and multiple local minima can complicate the optimization, while distinct Hamiltonians may produce nearly indistinguishable neutron-scattering responses.

Here we pursue a different strategy. Having constructed the forward map in Eq.~\eqref{eq:forward_map}, we evaluate it for a large set of Hamiltonians spanning the relevant parameter space and use the resulting dynamical responses to train a neural network to approximate the inverse relation. 
The computational effort associated with exploring the Hamiltonian parameter space is therefore shifted from the analysis of each individual experiment to a one-time training stage. 
Once trained, the network provides a direct estimate of the Hamiltonian parameters from the dynamical response. 
In the following section, we describe the construction of the training database and the neural-network models used to implement this inverse mapping.

\section{Neural networks}
\label{sec:NN}

We now implement the inverse-learning strategy introduced in Sec.~\ref{sec:inverse}. 
Using the exact forward solver developed in the previous section, we generate a synthetic database that associates each
set of Hamiltonian parameters $\{J_i\}$ with its corresponding zero-temperature dynamical spin response. 
The neural network is trained on this database to infer the Hamiltonian parameters directly from the spectral information.

For the fully polarized honeycomb models considered here, the dynamical response is completely characterized by the two magnon dispersions $\omega_{\pm}({\bf q})$ and their corresponding spectral weights $I_{\pm}({\bf q})$. 
We therefore use these four quantities, sampled from the region of the honeycomb-lattice Brillouin zone shown in Fig.~\ref{fig:honeycomb}(b), as input features for the neural networks, while the exchange parameters $J_1$--$J_8$ serve as the target outputs.

To examine how the representation of the spectral information affects Hamiltonian reconstruction, we consider three neural-network architectures: a fully connected neural network (FCNN), a one-dimensional convolutional neural network (CNN1D), and a two-dimensional convolutional neural network (CNN2D). 
These architectures process the same physical information using different representations of its momentum-space structure. 
In the following, we describe the generation and preprocessing of the training data, the three network architectures and their training procedure, and the metrics used to quantify the accuracy of the reconstructed Hamiltonian parameters.

\subsection{Data generation and description}
The input to the neural networks is constructed directly from the zero-temperature dynamical spin structure factor generated by the forward solver. 
For each Hamiltonian, we retain both components of the spectral information: the magnon dispersions $\omega_{\pm}({\bf q})$ and their corresponding spectral weights $I_{\pm}({\bf q})$ throughout the Brillouin zone. 
No additional descriptors or physically motivated combinations of these quantities are introduced. The neural networks are therefore trained to identify directly from the momentum-dependent spectral information the features that encode the underlying exchange interactions.

Including both the excitation energies and their spectral weights is important because they contain complementary information about the Hamiltonian. 
Whereas the magnon energies are determined by the eigenvalues of the spin-wave Hamiltonian, the spectral weights also depend on its eigenvectors and therefore contain information that is not generally
recoverable from the dispersion alone. 
As demonstrated in Sec.~\ref{sec:results}, this additional information plays an important role in resolving ambiguities between Hamiltonians that produce nearly identical magnon dispersions.

\subsubsection{Training data generation for neural networks}

To construct the training data for the supervised learning, all possible combinations of the eight Hamiltonian parameters, $J_1$--$J_8$, are first generated on a uniform grid. Since $J_1$ is chosen as the unit of energy, it is restricted to either -1 or 1. The remaining parameters, $J_2$--$J_8$, each take one of eight uniformly spaced values within the interval [-1.2, 1.2], inclusive of boundaries. This procedure generates a total of 4,194,304 unique Hamiltonian parameter combinations. Although realistic magnetic systems are expected to satisfy $J_2$--$J_8$ $\in$ [-1, 1], the slightly broader sampling range is adopted during training to reduce boundary effects when predicting physically relevant models.

Each of the 4,194,304 Hamiltonian parameter sets is passed to the forward solver described in Sec.~\ref{subsec:forward}. 
The solver evaluates the transverse dynamical structure factor on the uniform $16\times16$ momentum grid in the symmetry-reduced reciprocal-space domain shown in Fig.~\ref{fig:honeycomb}(b). 
At each of the 256 momenta, it returns the energies $\omega_{+}$ and $\omega_{-}$ of the upper and lower magnon branches, together with their respective spectral weights $I_{+}$ and $I_{-}$.

Each parameter set therefore corresponds to four momentum-dependent arrays, or $4\times256=1024$ spectral values. These parameter–response pairs define the synthetic dataset \verb|Honeycomb_Pure|. 
The spectral values serve as inputs to the neural networks, while the corresponding couplings $J_1$--$J_8$ serve as targets for supervised learning. 
The representations used by the individual network architectures are described in the following subsection.

Because the training dataset is several hundred gigabytes in size, an on-the-fly data generation strategy was adopted to reduce storage requirements and eliminate input/output (I/O) bottlenecks \cite{soler2026infinite}. Only the Hamiltonian parameters were stored and used for data splitting, while the corresponding magnon energies and scattering intensities were generated on the fly for each training batch. The training data were shuffled at every epoch, and the data generation process was parallelized to improve computational efficiency.

\subsubsection{Random test set generation for robustness evaluation}

Although the \verb|Honeycomb_Pure| dateset systematically samples the Hamiltonian parameter space through a structured grid of Hamiltonian parameter combinations, practical applications require the model to generalize to arbitrary Hamiltonian parameter combinations beyond this structured sample strategy. To better reflect such scenarios and further evaluate the generalization capability of the trained models, an independent \verb|Random_Pure| dataset was generated by randomly sampling the Hamiltonian parameters within physically realistic ranges.

Specifically, the Hamiltonian parameters $J_{2}$--$J_{8}$ were independently sampled from a uniform distribution over $[-1,\,1]$, while $J_{1}$ was assigned the values $+1$ and $-1$ with equal probability, yielding 10,000 samples for each class. The sampled Hamiltonian parameter combinations were then processed using the same forward solver employed for the \verb|Honeycomb_Pure|, and each resulting neutron scattering spectrum was represented by the same 1024-dimensional feature vector.

\subsection{Data preprocessing}

\subsubsection{Noise injection}

The \verb|Honeycomb_Pure| and \verb|Random_Pure| datasets were initially generated as clean synthetic datasets. However, real neutron scattering experiments are inevitably contaminated by measurement noise arising from instrumental limitations and experimental uncertainty. To evaluate the robustness of the trained models, artificial Gaussian noise was subsequently added to both datasets to simulate experimental uncertainty \cite{tsiligkaridis2023diverse}. This represents a standard approach to testing and regularizing neural networks. We will further discuss how neutron counting noise and other information obscuring processes associated with a real experiment can be modeled more precisely in Sec.~\ref{sec:closing_loop}.

For each sample, independent Gaussian noise was generated from a standard normal distribution and then scaled by two parameters, $\sigma_{\omega}$ and $\sigma_{I}$, which control the noise amplitude of the dispersion and intensity features, respectively. For the dispersion features, the noise standard deviation is defined as $\sigma_{\omega}$ multiplied by the maximum bandwidth of the upper and lower magnon branches, where the bandwidth is computed as the difference between the maximum and minimum dispersion values within each sample. For the scattering intensity features, the noise standard deviation is defined as $\sigma_{I}$ multiplied by the maximum intensity across both magnon branches. Scaling the noise by the characteristic energy and intensity of each spectrum provides a realistic representation of experimental measurement uncertainty while ensuring that the noise remains proportional to the scale of each spectrum. 

Two data sets were generated in this fashion, \verb+Honeycomb+ and \verb+Random+, corresponding respectively to the \verb|Honeycomb_Pure| (parameters sampled on a regular grid) and \verb|Random_pure| (uniformly sampled parameters in $[-1, 1]$) datasets. Unless otherwise specified, the same noise level is applied to both the dispersion and intensity features, i.e., $\sigma_{\omega}=\sigma_{I}$. To evaluate the robustness of the proposed models under increasing experimental uncertainty, three noise amplitudes, 0, 0.05, and 0.10,  were considered throughout this work.

\subsubsection{Dataset splitting}

The \verb|Honeycomb| dataset was randomly partitioned into Training, Validation, and Test sets using an 8:1:1 ratio, corresponding to 80\%, 10\%, and 10\% of the total samples, respectively. The Training and Validataion sets were used for neural network training and model tuning, whereas the Test and independent Random sets were reserved exclusively for evaluating predictive performance and robustness under increasing noise amplitude \cite{babaei2025impact, kahloot2021algorithmic}.

\subsubsection{Features normalization}

Following data partitioning, z-score normalization was applied to the input features. The feature-wise mean and standard deviation were computed from the Training set and subsequently used to standardize the Training, Validation, Test, and Random sets. As a result, the normalized input features have approximately zero mean and unit variance. This preprocessing improves training stability, facilitates numerical optimization, and accelerates model convergence \cite{kim2025investigating}.

\subsubsection{Input representations}

The normalized spectral features were then reshaped according to the input requirements of the three neural network architectures \cite{goodfellow2016deep}. For the CNN-based models, the magnon dispersion branches ($\omega_{+}$ and $\omega_{-}$) and neutron scattering intensity maps ($I_{+}$ and $I_{-}$) were represented as four spectral channels. Each channel was represented as a one-dimensional sequence of 256 momentum points for CNN1D and retained its original $16 \times 16$ momentum-grid structure for CNN2D. The resulting input representations are summarized in Table~\ref{tab:input_representation}, where $m$ denotes the number of input samples.

\begin{table}[htbp]
    \centering
    \caption{Input representations of the neural networks.}
    \label{tab:input_representation}
    
    \begin{tabular}{lcl}
    \toprule
    \textbf{Model} & \textbf{Input Shape} & \textbf{Description} \\
    \midrule
    FCNN  & $(m,1024)$      & Flattened feature vector \\
    CNN1D & $(m,4,256)$     & Four-channel 1D representation \\
    CNN2D & $(m,4,16,16)$   & Four-channel 2D representation \\
    \bottomrule
\end{tabular}

\end{table}

\subsubsection{Target processing}

As mentioned, the Hamiltonian parameters $J_{1}$--$J_{8}$ serve as the target outputs of the neural network models. Since $J_{1}$ is restricted to the discrete values of $-1$ and $1$, it is encoded as class labels $0$ and $1$, respectively, to satisfy the input requirements of the PyTorch cross-entropy loss function, which expects nonnegative integer class indices \cite{pointer2019programming}. The remaining parameters, $J_{2}$--$J_{8}$, are retained in their original continuous form as regression targets.

\subsection{Metrics}

The predictive performance of the proposed models was evaluated separately for the classification and regression outputs. Classification accuracy was used to evaluate the binary prediction of $J_{1}$ \cite{hastie2009elements}, while the relative error (RE) was used to evaluate the continuous predictions of $J_{2}$--$J_{8}$.

The classification accuracy is defined as:

\begin{equation}
    \text{Accuracy} =
    \left(
    \frac{N_{\mathrm{correct}}}
    {N_{\mathrm{total}}}
    \right)
    \times 100\%,
\end{equation}
where $N_{\mathrm{correct}}$ represents the number of correctly classified samples, and $N_{\mathrm{total}}$ represents the total number of samples in the evaluated dataset.

The relative error RE of the $n^{\text{th}}$ exchange interaction, $J_n$ is defined as:
\begin{equation}
    \mathrm{RE}_n
    =
    \frac{1}{m}
    \sum_{i=1}^{m}
    \frac{\left| J_n^i - \hat{J}_n^i \right|}
    {\left| J_n^i \right|}
    \left(
    1 - \tanh\left(\sigma\frac{\left|J_1^i\right|}{\left| J_n^i \right|}\right)
    \right),
\end{equation}
where $J_n^i$ and $\hat{J}_n^i$ respectively represent the true and predicted values of a target parameter, $J_n$,   and $m$ is the total number of evaluated samples. The scaling parameter $\sigma$ is defined as  $\sigma = 0.02$.  The weighting factor  $1-\tanh(\sigma |J_1^i|/|J_n^i|)$ is introduced to prevent disproportionately large relative error when the true parameter value is close to zero. For each sample, the average RE is computed by averaging the relative errors over the seven regression targets, $J_2$--$J_8$. Consequently, higher classification accuracy and lower relative error values indicate more accurate Hamiltonian reconstructions.

\subsection{Neural network configuration and training settings}

Following data preprocessing and evaluation metrics definition, the three neural network architectures were trained using the prepared datasets: a fully connected neural network (FCNN), a one-dimensional convolutional neural network (CNN1D), and a two-dimensional convolutional neural network (CNN2D). All models were implemented using the PyTorch deep learning framework \cite{ketkar2021introduction}.

The three neural network architectures were selected to compare different strategies for reconstructing Hamiltonian parameters from neutron scattering spectra. The FCNN treats the input spectrum as a flattened feature vector without explicitly modeling local relationships among neighboring spectral features \cite{goodfellow2016deep, sazli2006brief}. In contrast, the CNN1D processes the input as a one-dimensional sequence sampled along the momentum path, enabling the network to progressively exploit local spectral correlations through hierarchical feature extraction \cite{goodfellow2016deep, kiranyaz20211d}. The CNN2D preserves the two-dimensional geometry of reciprocal space, allowing the network to increasingly learn both local patterns and spatial relationships \cite{goodfellow2016deep, gramlich2026convolutional}. Comparing these architectures provide insights into the trade-offs among feature extraction capability, model complexity, parameter efficiency, and prediction accuracy, offering practical guidance for selecting an appropriate architecture for different application scenarios.

Despite their different backbone architectures, all three models adopt the same multi-task learning framework to simultaneously predict the binary parameter $J_{1}$ and the continuous parameters $J_{2}$--$J_{8}$ \cite{crawshaw2020multi, ruder2017overview}. For each model, it consists of a shared backbone network for feature extraction, followed by two task-specific output heads: a classification head with two output neurons for predicting $J_{1}$, and a regression head with seven output neurons for predicting $J_{2}$--$J_{8}$. Consequently, the three models differ only in the design of the backbone networks, while the output heads are identical across all architectures. The backbone architectures are described below.

\subsubsection{Fully connected neural network (FCNN)}

\begin{figure}
    \centering
    \includegraphics[width=0.48\textwidth]{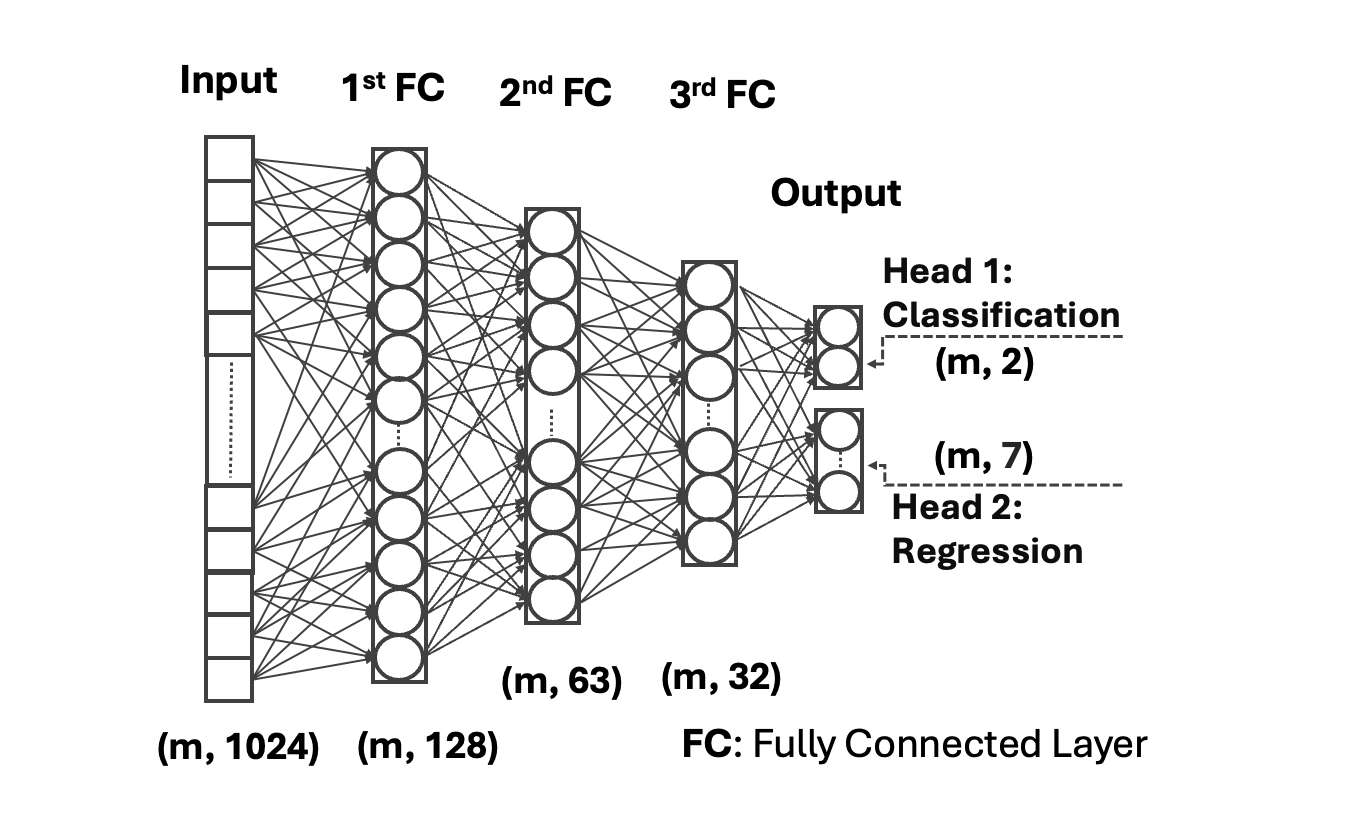}
    \caption{Architecture of the FCNN model.}
    \label{fig:FCNN_architecture}
\end{figure}

As shown in Figure~\ref{fig:FCNN_architecture}, the FCNN consists of an input layer, three fully connected hidden layers, and the common multi-task output layer described above. The network contains 141,767 trainable parameters and takes a flattened 1024-dimensional feature vector as input. It learns global feature representations through three hidden layers containing 128, 64, and 32 neurons, respectively, each followed by a ReLU activation function \cite{agarap2018deep, parhi2020role}. The extracted features are then fed into the common multi-task output layers for simultaneous classification of $J_{1}$ and regression of $J_{2}$--$J_{8}$. The FCNN contains 141,767 trainable parameters.

\subsubsection{One-dimensional convolutional neural network (CNN1D)}

\begin{figure}
    \centering
    \includegraphics[width=0.48\textwidth]{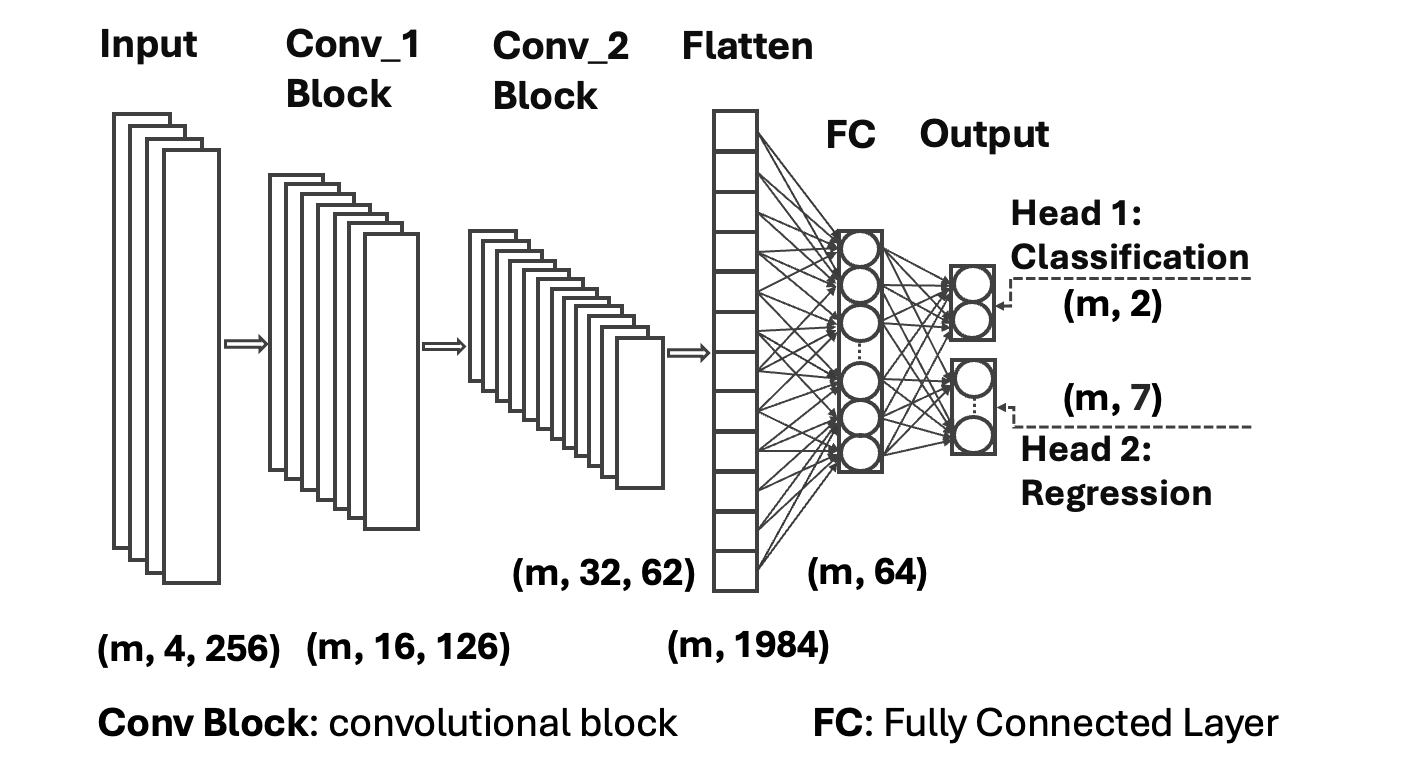}
    \caption{Architecture of the CNN1D model.}
    \label{fig:cnn1d_architecture}
\end{figure}

Figure~\ref{fig:cnn1d_architecture} illustrate the CNN1D architecture, which consists of one input layer, two convolutional blocks, a fully connected layer, and the common multi-task output layer. Each convolutional block comprises a one-dimensional convolutional layer, a ReLU activation function, and a max-pooling operation \cite{kiranyaz20211d}.

The first convolutional layer employs 16 kernels with a kernel size of 5 and a stride of 1, while the second convolutional layer uses 32 kernels with a kernel size of 3 and a stride of 1. Following each convolutional layer, the same max-pooling operation with a kernel size of 2 and a stride of 2 is applied to reduce the feature dimensionality while preserving the most informative features \cite{goodfellow2016deep, zafar2022comparison}. The resulting feature maps are then flattened and passed to a fully connected layer containing 64 neurons with ReLU activation before being fed into the common multi-task output layer. The CNN1D model contains 129,399 trainable parameters.

\subsubsection{Two-dimensional convolutional neural network (CNN2D)}

\begin{figure}
    \centering
    \includegraphics[width=0.48\textwidth]{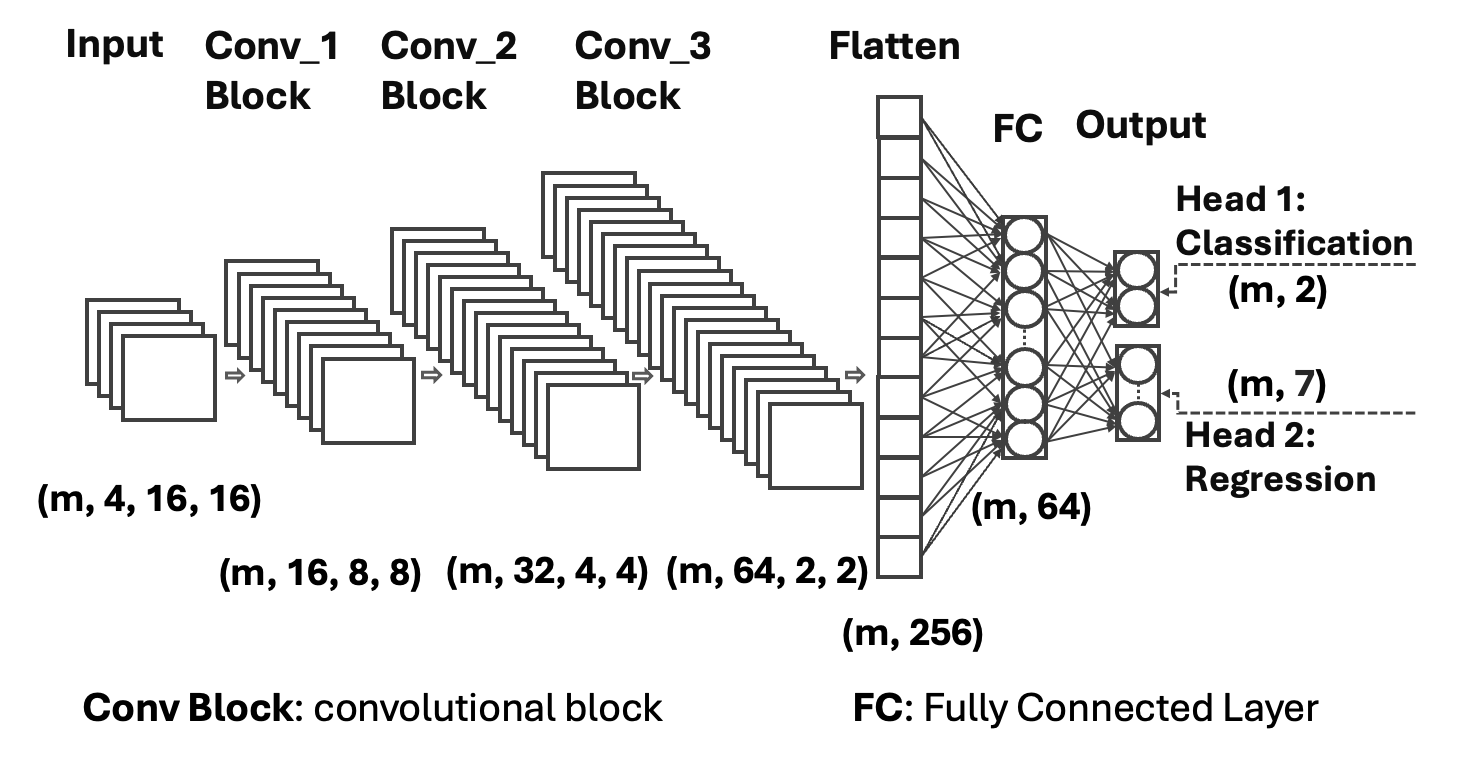}
    \caption{Architecture of the CNN2D model.}
    \label{fig:cnn2d_architecture}
\end{figure}

The CNN2D architecture, as shown in Figure~\ref{fig:cnn2d_architecture}, consists of one input layer, three convolutional blocks, a fully connected layer, and the multi-task output layer. Each convolutional block comprises a two-dimensional convolutional layer followed by a ReLU activation function, and a max-pooling operation \cite{habibi2017convolutional, saleem2022comparative}.

The three convolutional layers employ 16, 32, and 64 kernels, respectively, each using a $3 \times 3$ kernel with a stride of 1. Before each convolution, zero-padding is applied to preserve the spatial dimensions of the feature maps \cite{yu2023efficient, han2023deep}. Each convolution is followed by a max-pooling operation with a $2 \times 2$ kernel and a stride of 2. After the final convolutional block, the feature maps are flattened and passed to a fully connected layer containing 64 neurons with ReLU activation before being fed into the common multi-task output layer for simultaneous classification of $J_{1}$ and regression of $J_{2}$--$J_{8}$. The CNN2D model contains 38,451 trainable parameters.

To ensure a fair comparison, all three neural networks were trained using the same optimization settings. The training objective was a multi-task loss combining Cross-Entropy loss for the classification task \cite{liu2016large, kodamanchili2026training} and Mean Squared Error (MSE) loss for the regression task \cite{wang2024regression}. Model parameters were optimized using the Adam optimizer with a learning rate of $1 \times 10^{-4}$ \cite{dereich2024convergence, bock2018improvement}. Training was conducted for 150 epochs with a batch size of 32.  Training and validation losses were recorded at each epoch and  used to generate the learning curves presented in the Results section.

\section{Results}
\label{sec:results}

\subsection{Information content and invertibility}
\label{subsec:invertibility}

An important feature of the supervised-learning framework is that the training process itself provides a diagnostic of the information content of the neutron-scattering data. 
Reliable Hamiltonian reconstruction requires that the observables supplied to the neural network contain sufficient information to distinguish different Hamiltonians within the prescribed model family. 
If distinct Hamiltonians produce indistinguishable inputs, the inverse problem is underconstrained and no inference method can uniquely reconstruct the microscopic parameters. 
Systematic failures of the reconstruction can therefore reveal that the chosen set of observables does not contain enough information to determine the Hamiltonian.

We illustrate this principle by first asking whether the magnon dispersions alone are sufficient for Hamiltonian reconstruction. 
As discussed in Sec.~\ref{subsec:forward}, the dispersions $\omega_{\pm}({\bf q})$ are determined by the eigenvalues of the spin-wave Hamiltonian. 
To assess the information contained in the excitation energies, we first consider a reduced three-parameter problem involving $J_1$, $J_2$, and $J_3$ and train a CNN1D model using only $\omega_{\pm}(q)$ as input.
The dataset was generated following the same procedure as the Honeycomb dataset, except that a linear density of 40 was used.
A noise level of $\sigma=0.05$ was applied to both the training and evaluation.

\begin{figure}
    \centering
    \includegraphics[width=0.48\textwidth]{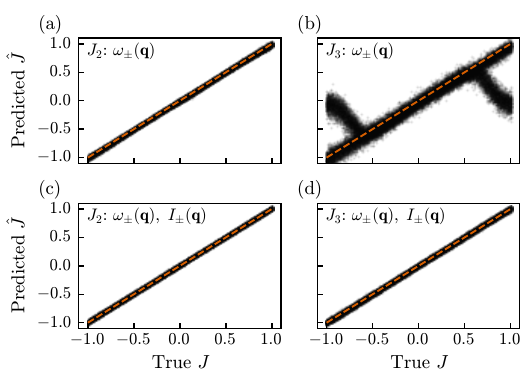}
    \caption{Hamiltonian reconstruction using (a)-(b) the magnon dispersions $\omega_{\pm}({\bf q})$
    alone and (c)-(d) the magnon dispersions $\omega_{\pm}({\bf q})$ together with their spectral
    weights $I_{\pm}({\bf q})$. The outliers obtained from the dispersion alone reveal
    regions in which the available spectral information does not uniquely
    constrain the Hamiltonian. Including the spectral weights largely
    removes these ambiguities.}
    \label{fig:invertibility}
\end{figure}

As shown in Fig.~\ref{fig:invertibility}(a) and Fig.~\ref{fig:invertibility}(b), for a randomly sampled three-parameter dataset generated using the same procedure as the \verb|Random| set, the prediction for $J_2$ closely match the true values, whereas those for $J_3$ exhibit distinct branches that deviate substantially from the identity line. Examination of these cases reveals that distinct sets of Hamiltonian parameters can produce nearly indistinguishable magnon dispersions. The resulting ambiguity is therefore not simply a limitation of the neural network: the information supplied to the network is insufficient to uniquely constrain the Hamiltonian in these regions of parameter space.

The dynamical spin structure factor, however, contains information beyond the excitation energies. 
In particular, the spectral weights $I_{\pm}({\bf q})$ also depend on the eigenvectors of the spin-wave Hamiltonian and therefore provide information about the microscopic interactions that is not generally contained in the magnon dispersions.
We consequently repeat the reconstruction using both $\omega_{\pm}({\bf q})$ and $I_{\pm}({\bf q})$ as inputs. 
As shown in Fig.~\ref{fig:invertibility}(c) and Fig.~\ref{fig:invertibility}(d), the previously observed outliers are significantly eliminated, and the overall average relative prediction error of $J_2$ and $J_3$ decreases from $18.37\%$ to $4.04\%$, and accuracy of $J_1$ increases from $98.8\%$ to $100\%$. 
The improvement demonstrates that the spectral weights provide an independent constraint that resolves ambiguities left by the excitation energies alone.

The same ambiguity persists in the full eight-parameter model family. Equation~\eqref{eq:magnon_dispersion} shows that two parameter sets at the same field produce identical magnon dispersions throughout the Brillouin zone if their same-sublattice couplings \(J_2\), \(J_5\), and \(J_6\) are identical and their opposite-sublattice Fourier sums satisfy
\begin{equation}
\begin{aligned}
J^{AB\prime}({\bf 0}) &= J^{AB}({\bf 0}),\
|J^{AB\prime}({\bf q})| &= |J^{AB}({\bf q})|
\quad \forall{\bf q}.
\end{aligned}
\label{eq:same_disp_condition}
\end{equation}
Here, the prime labels the second parameter set. Importantly, these conditions constrain only the magnitude of \(J^{AB}({\bf q})\), not its phase. Distinct Hamiltonians can therefore have identical magnon energies but different spectral weights.

\begin{figure}
    \centering
    \includegraphics[width=.4\textwidth]{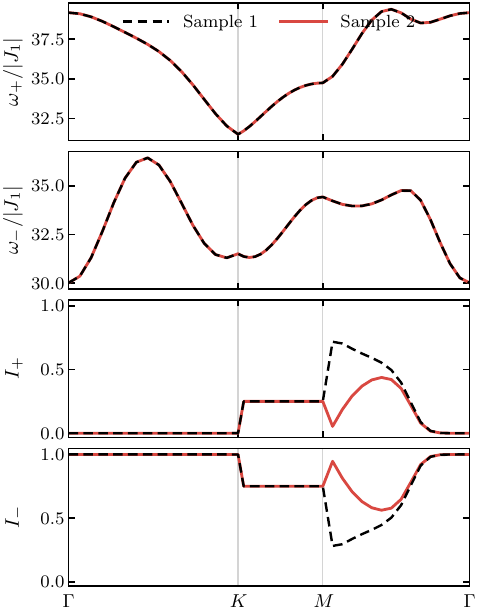}
    \caption{Magnon dispersions $\omega_\pm/|J_1|$ and intensities $I_\pm$ along the momentum path $\Gamma$--$K$--$M$--$\Gamma$ for two distinct eight-parameter Hamiltonians. The dispersions are indistinguishable at the plotted resolution along the entire path, whereas the intensities differ most visibly along the $M$--$\Gamma$ segment. With $a = 0.8571$, $b = 0.1714$: 
    Sample 1: $\mathbf{J} = (-1, a, b, -a, -a, -b, b, -a)$;
    Sample 2 (red solid lines): $\mathbf{J} = (-1, a, -a, b, -a, -b, -a, b)$. 
    The two parameter sets satisfy the conditions in Eq.~\eqref{eq:same_disp_condition}.
}
    \label{fig:similar_omega_different_intensity}
\end{figure}

Figure~\ref{fig:similar_omega_different_intensity} illustrates this ambiguity with two distinct Hamiltonians whose magnon dispersions are nearly indistinguishable but whose spectral-weight distributions differ substantially. 
Excitation energies alone do not uniquely distinguish these models, whereas their spectral weights resolve the ambiguity.

This example illustrates a broader role of supervised learning in the inverse neutron-scattering problem. 
In addition to reconstructing Hamiltonian parameters, the training process provides a systematic way to determine whether a proposed set of experimentally accessible observables contains sufficient information to constrain those parameters.
Persistent reconstruction ambiguities signal that the inverse problem is underconstrained and identify the need for additional experimental information. 
In the present example, the magnon dispersions alone are insufficient in parts of parameter space, while the spectral weights
provide the missing information. 
More generally, the same strategy can be used to determine the minimal type of neutron-scattering data required for reliable Hamiltonian reconstruction within a given model family.
This information can then be used to guide experiment design by identifying, before measurements are performed, which observables must be measured to determine the microscopic Hamiltonian. 
For this reason, both $\omega_{\pm}({\bf q})$ and $I_{\pm}({\bf q})$ are included in all subsequent eight-parameter reconstructions.

\subsection{Accuracy and robustness of Hamiltonian reconstruction}
\label{subsec:accuracy_robustness}

\begin{figure}
    \centering
    \includegraphics[width=0.45\textwidth]{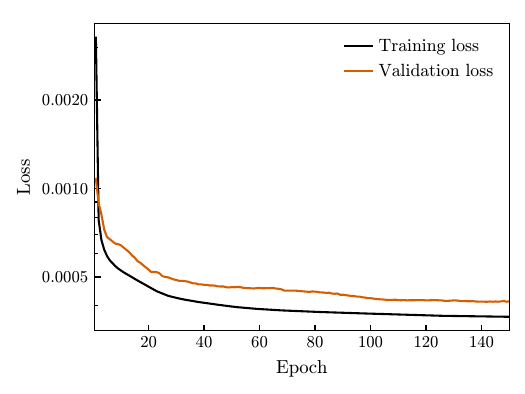}
    \caption{Learning curves for the CNN1D model with noise set to $\sigma = 0.05$. The combined training and validation losses are shown as functions of the training epoch.}
    \label{fig:CNN1D_learning_curve}
\end{figure}

Having established that the combined magnon dispersions and spectral
weights provide sufficient information to constrain the inverse problem,
we now examine how accurately the neural network reconstructs the full
eight-parameter Hamiltonian and how robust this reconstruction remains
when the input spectra are contaminated by noise. We focus first on the
CNN1D architecture and consider two complementary evaluation sets. The
Test set is drawn from the same structured parameter grid used to construct
the training database, whereas the Random set consists of independently
sampled Hamiltonians that do not belong to this grid. 
The latter therefore provides a direct test of the ability of the network 
to interpolate between the Hamiltonians used during training.

Figure~\ref{fig:CNN1D_learning_curve} shows the learning curves of the CNN1D model for \(\sigma=0.05\). The training and validation losses decrease rapidly during the initial epochs and subsequently converge to stable values. The validation loss closely tracks the training loss throughout optimization, demonstrating stable convergence.

\begin{figure}
    \centering
    \includegraphics[width=0.45\textwidth]{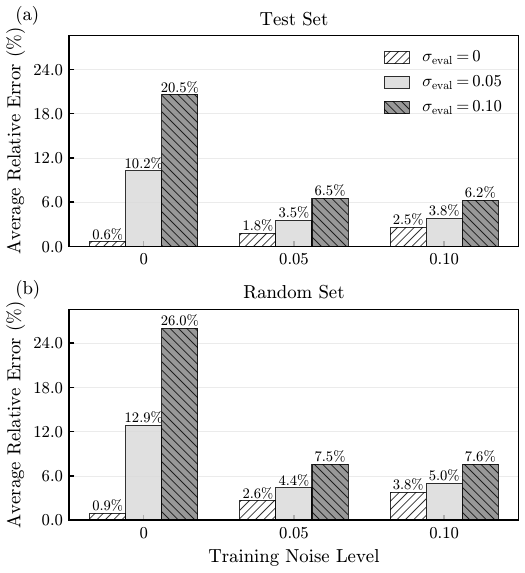}
    \caption{Average relative error of the CNN1D model on (a) the Test set
    and (b) the independently sampled Random set. Models are trained with
    noise amplitudes $\sigma=0$, $0.05$, and $0.10$ and evaluated at the
    same three noise levels.}
    \label{fig:test_set_random_set_RE_panels}
\end{figure}

Figure~\ref{fig:test_set_random_set_RE_panels} shows the average relative
error for CNN1D models trained with noise amplitudes $\sigma=0$, $0.05$,
and $0.10$ and evaluated at the same three noise levels. In the absence of
noise, the model trained on exact spectra reconstructs the Hamiltonian with
high accuracy: the average relative error on the Test set is only $0.6\%$.
The error remains small on the independently sampled Random set, showing
that the network is not simply reproducing Hamiltonians contained in the
training grid but has learned to interpolate the inverse relation throughout
the sampled parameter space.

We next examine the stability of the reconstruction against uncertainty in
the spectral data. For all three training configurations, the prediction
error increases continuously as the noise applied to the input spectra is
increased. For example, on the Test set the model trained on noise-free
spectra exhibits average relative errors of $0.6\%$, $10.2\%$, and
$20.5\%$ when evaluated at $\sigma=0$, $0.05$, and $0.10$, respectively.
The same qualitative behavior is observed on the Random set, with somewhat
larger errors because these Hamiltonians are not drawn from the structured
training grid. Importantly, the reconstruction degrades progressively with
increasing noise rather than exhibiting an abrupt loss of predictive
capability.

The results also show that robustness can be improved by incorporating the
expected uncertainty directly into the training data \cite{bishop1995training, an1996effects}. For noise-free input,
training on exact spectra gives the highest accuracy. At finite evaluation
noise, however, networks trained with noisy spectra perform substantially
better. The model trained with $\sigma=0.05$ provides the best compromise
between accuracy on clean data and robustness to moderate noise, whereas
the model trained with $\sigma=0.10$ performs best when evaluated at the
largest noise level. This demonstrates that the synthetic training database
can be adapted to the expected experimental uncertainty, allowing the
learned inverse relation to be optimized for the quality of the available
data. In the remainder of this work, unless otherwise specified, we use
$\sigma=0.05$ for both training and evaluation.

\begin{figure*}
    \centering
    \includegraphics[width=.98\textwidth]{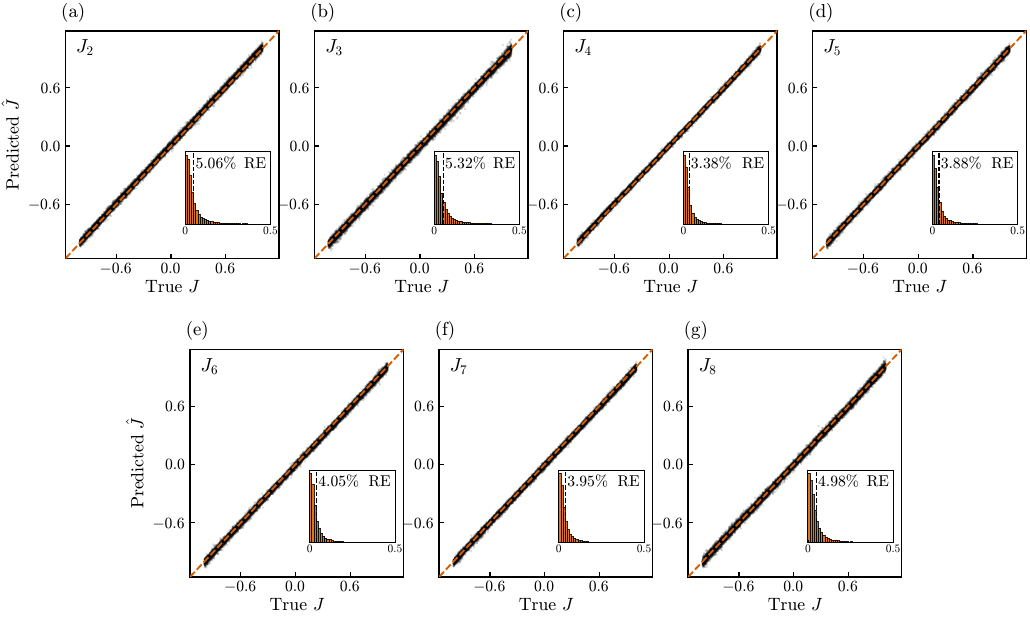}
    \caption{Predicted versus true values of the continuous Hamiltonian parameters
    $J_2$--$J_8$, reconstructed by the CNN1D model on the Random set at
    $\sigma = 0.05$. Panels (a)--(g) show $J_2$ through $J_8$; the orange
    dashed line is the identity line $\hat{J} = J$, indicating perfect
    agreement between predicted and true values. The inset in each panel
    shows the corresponding scaled-relative-error (RE) distribution, with
    the dashed vertical line and adjacent value marking its mean.}
    \label{fig:CNN1D_Js_with_error_insets}
\end{figure*}

To resolve the reconstruction accuracy at the level of individual
interactions, we examine the CNN1D predictions on the Random set at
$\sigma=0.05$. The insets of Fig.~\ref{fig:CNN1D_Js_with_error_insets} show the relative-error
distributions for the seven continuous parameters $J_2$--$J_8$. The
distributions are strongly concentrated at small errors, with average
relative errors ranging from $3.38\%$ to $5.32\%$. Only $J_2$ and $J_3$
slightly exceed $5\%$, with average relative errors of $5.06\%$ and
$5.32\%$, respectively, while the remaining five parameters are
reconstructed with average errors below $5\%$.

The corresponding predicted-versus-true values are shown in
Fig.~\ref{fig:CNN1D_Js_with_error_insets}. The predictions remain closely clustered
around the identity line throughout the parameter range for all seven
continuous couplings. Together with the $100\%$ classification accuracy
obtained for the sign of $J_1$, these results demonstrate that the network
can simultaneously reconstruct all eight Hamiltonian parameters for
previously unseen Hamiltonians while remaining robust to moderate
uncertainty in the spectral data.

%\begin{figure}
%    \centering
%    \includegraphics[width=0.48\textwidth]{paper_plots/CNN1D_Js_RE_hist.pdf}
%    \caption{Relative-error distributions for the reconstructed continuous
%    Hamiltonian parameters $J_2$--$J_8$ on the Random set at
%    $\sigma=0.05$. \takwi{Combine fig. 8 to fig. 9 as insets?}}
%    \label{fig:CNN1D_Js_RE_hist}
%\end{figure}

%\begin{figure}
%    \centering
%    \includegraphics[width=0.48\textwidth]{paper_plots/CNN1D_Js_scatter.png}
%    \caption{Predicted versus true values of the continuous Hamiltonian parameters $J_2$--$J_8$ for the CNN1D model evaluated on the Random set at $\sigma = 0.05$. The red dashed identity line indicates perfect agreement between the predicted and true values.}
%    \label{fig:CNN1D_Js_scatter}
%\end{figure}

\subsection{Comparison of neural-network architectures}
\label{subsec:architecture_comparison}

We finally ask to what extent the reconstruction depends on the particular
neural-network architecture. The FCNN, CNN1D, and CNN2D models process the
same spectral information using substantially different representations:
the FCNN treats the spectrum as a flattened vector, the CNN1D exploits
local correlations in a one-dimensional ordering of momentum points, and
the CNN2D explicitly preserves the two-dimensional geometry of reciprocal
space. A strong dependence on architecture would indicate that successful
reconstruction relies on a particular inductive bias, whereas comparable
performance would suggest that the inverse relation can be learned robustly
from the spectral information itself.

\begin{figure}[H]
    \centering
    \includegraphics[width=0.45\textwidth]{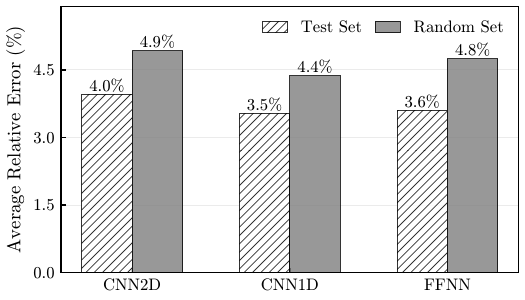}
    \caption{Average relative errors of the FCNN, CNN1D, and CNN2D models
    on the Test and Random sets for training and evaluation noise
    $\sigma=0.05$.}
    \label{fig:three_models_RE_comparison}
\end{figure}

Figure~\ref{fig:three_models_RE_comparison} compares the three architectures
on the Test and Random sets using $\sigma=0.05$ for both training and
evaluation. All three models achieve $100\%$ classification accuracy for
$J_1$. Their regression accuracies for $J_2$--$J_8$ are also remarkably
similar. On the Test set, the average relative errors range from $3.53\%$
to $3.96\%$, while on the independently sampled Random set they range from
$4.37\%$ to $4.93\%$. The modest increase on the Random set is common to
all three architectures and reflects the more stringent interpolation test
provided by Hamiltonians that do not belong to the structured training
grid.

The comparable performance of the three architectures is itself an
important result. Despite their different representations and inductive
biases, each network is able to learn the inverse relation with similar
accuracy. The successful reconstruction is therefore not tied to a
particular neural-network architecture, but instead reflects the
information contained in the dynamical response and the learnability of
the inverse map within the model family considered here.

The architectures nevertheless differ substantially in computational
complexity, as summarized in Table~\ref{tab:model_comparison}. In
particular, CNN2D achieves comparable reconstruction accuracy with only
38,451 trainable parameters, compared with 141,767 for FCNN and 129,399
for CNN1D. Conversely, the simpler FCNN has the shortest training time.
Once trained, however, all three models perform inference at rates of
approximately $2\times10^4$ spectra per second, making the computational
cost of Hamiltonian reconstruction negligible compared with the generation
of the training database.  The reported computation times throughout this work were measured on a Mac Pro equipped with a 3.0 GHz 10-core Intel Xeon W processor.

\begin{table}[htbp]
\centering
\caption{Model complexity, training time, and inference speed for the three
neural-network architectures.}
\label{tab:model_comparison}

\begin{tabular}{lccc}
\toprule
\textbf{Model} &
\textbf{Parameters} &
\textbf{Training Time} &
\textbf{Inference Speed}\\
\midrule
FCNN  & 141,767 & 15.25 h & 21986 samples/s\\
CNN1D & 129,399 & 31.71 h & 18767 samples/s\\
CNN2D &  38,451 & 40.49 h & 18892 samples/s\\
\bottomrule
\end{tabular}
\end{table}

\begin{table*}[htbp]
    \centering
    \small
    \caption{Dependence of the reconstruction accuracy and computational
    cost on the number of exchange parameters $N_p$ and the number $b$ of
    grid points per continuous parameter direction. The grid spacing is
    $\Delta J=2.4/(b-1)$, and the total number of models is
    $N_{\mathrm{grid}}=2b^{N_p-1}$. The quantity
    $\overline{\mathrm{RE}}_{\mathrm{R}}$ is the average relative error
    over the continuous parameters on the independent Random set, while
    $\mathrm{RE}_{\max,\mathrm{R}}$ is the largest parameter-resolved
    average relative error.Training times correspond to 150 epochs on the hardware used throughout
    this work. The
    classification accuracy for $J_1$ is $100\%$ in all cases.}
    \label{tab:grid_scaling}
    \begin{tabular}{cccccccc}
    \hline\hline
    $N_p$ & $b$ & $h$ & $N_{\mathrm{grid}}$
    & $\overline{\mathrm{RE}}_{\mathrm{R}}$
    & $\mathrm{RE}_{\max,\mathrm{R}}$
    & $t_{\mathrm{train}}$ (h)
    & Inference rate (spectra/s) \\
    \hline
    3 & 20 & 10 & 800
    & $4.92\%$ & $5.65\%\;(J_3)$
    & 0.0103 & 18,522 \\
    3 & 30 & 10 & 1,800
    & $4.76\%$ & $6.11\%\;(J_3)$
    & 0.0150 & 18,396 \\
    3 & 40 & 10 & 3,200
    & $4.04\%$ & $4.58\%\;(J_3)$
    & 0.0228 & 18,442 \\
    6 & 8 & 10 & 65,536
    & $4.51\%$ & $6.94\%\;(J_3)$
    & 0.479 & 18,791 \\
    6 & 10 & 10 & 200,000
    & $4.21\%$ & $6.52\%\;(J_3)$
    & 1.79 & 18,523 \\
    8 & 8 & 30 & 4,194,304
    & $4.89\%$ & $5.32\%\;(J_3)$
    & 31.7 & 18,767 \\
    \hline\hline
    \end{tabular}
\end{table*}

\subsection{Scaling with the dimension of parameter space}

A potential limitation of the present approach is the size of the
synthetic training database. Let $N_p$ denote the number of exchange
parameters and $b$ the number of grid points along each continuous
parameter direction. Because $J_1$ is restricted to the two discrete
values $J_1=\pm1$, while $J_2,\ldots,J_{N_p}$ are sampled continuously,
the dimension of the continuous parameter space is $d=N_p-1$. For the
interval $J_i\in[-1.2,1.2]$ used to generate the training set, $b$
uniformly spaced points correspond to a grid spacing
\begin{equation}
\Delta J=\frac{2.4}{b-1}.
\label{eq:grid_spacing}
\end{equation}
Equivalently, the linear sampling density is
$\rho_{\mathrm{grid}}=1/\Delta J$. The total number of Hamiltonians in
the structured grid is
\begin{equation}
N_{\mathrm{grid}}=2b^{N_p-1}.
\label{eq:grid_size}
\end{equation}
Thus, at fixed $b$, the size of the training set grows exponentially
with the number of continuous parameters, as expected from the curse
of dimensionality.

To investigate how the required number of points per direction changes
with the dimension of the model space, we repeated the CNN1D
reconstruction for model families containing three, six, and eight
exchange parameters. In all cases, the spectra were evaluated on the
same $16\times16$ momentum grid, and a noise amplitude $\sigma=0.05$
was used for both training and evaluation. The reconstruction accuracy
was evaluated using independently and uniformly sampled Hamiltonians
that do not belong to the structured training grid. The results are
summarized in Table~\ref{tab:grid_scaling}.

For the three-parameter model, $b=40$ is the smallest of the tested
values for which the average relative error of every continuous
coupling remains below $5\%$. By contrast, $b=8$--$10$ already yields
average errors near the target accuracy for the six- and
eight-parameter models. For the six-parameter model, the overall
average errors are $4.51\%$ and $4.21\%$ for $b=8$ and $b=10$,
respectively, although the error in $J_3$ remains above $6\%$. For the
eight-parameter model, $b=8$ gives an overall average error of $4.89\%$;
only $J_2$ and $J_3$ lie slightly above $5\%$, with errors of $5.06\%$
and $5.32\%$, respectively. Although the available data do not determine
the precise minimum value of $b$ for every $N_p$, they demonstrate that
the number of grid points required along each parameter direction to
obtain approximately $5\%$ reconstruction accuracy decreases
substantially as the dimension of the model space increases.

A simple geometric consideration provides a possible explanation for
this trend. The coordination number of a $d$-dimensional hypercubic
grid is $2d$, but the relevant quantity for interpolation is the number
of vertices of the elementary hypercube containing a new model. A
generic Hamiltonian that does not belong to the training grid lies
inside an elementary $d$-dimensional hypercube bounded by $2^d$ grid
vertices. Consequently, as $d$ increases, each elementary cell is
surrounded by an exponentially growing number of training samples,
even though the number of points along each individual direction can
be reduced. If the inverse map from the dynamical response to the
Hamiltonian parameters is sufficiently smooth, these corner samples
provide multiple correlated constraints on the interpolation within
the cell. This geometric argument should be regarded as a heuristic
rather than a general error bound, because the required value of $b$
also depends on the smoothness and anisotropy of the inverse map and
on the information content of the spectra.

The reduction in the required linear grid size does not eliminate the
exponential growth in Eq.~\eqref{eq:grid_size}, but it mitigates it
substantially. For example, retaining the three-parameter value $b=40$
for the eight-parameter model would require
\begin{equation}
2\times40^7 \simeq 3.28\times10^{11}
\end{equation}
spectra. By contrast, using $b=8$ reduces the database to approximately
$4.2\times10^6$ spectra while preserving parameter-resolved errors
close to $5\%$. The computational cost of training scales approximately linearly with
the number of models in the grid. For the present implementation, the
measured training times in Table~\ref{tab:grid_scaling} are well
described by the practical estimate
\begin{equation}
t_{\rm train}\simeq
7.6\,{\rm h}\left(\frac{N_{\rm grid}}{10^6}\right),
\label{eq:training_time}
\end{equation}
for 150 training epochs on the hardware used in this work. Accordingly,
training on the full eight-parameter grid containing approximately
$4.2\times10^6$ models requires $31.7$ hours. By contrast, the inference
speed is nearly independent of the size and dimension of the training
set, remaining close to $1.85\times10^4$ spectra per second. Thus, the
computational cost is concentrated in the one-time training stage,
whereas the subsequent reconstruction of Hamiltonian parameters is
essentially instantaneous.

The reduction in the required number of points per
direction is therefore essential for making high-dimensional
Hamiltonian reconstruction computationally practical. The present
results further suggest that model families containing up to
approximately ten parameters may remain accessible. Such extensions
are particularly relevant to metallic magnets, where the long-ranged
and oscillatory character of RKKY interactions often requires the
inclusion of exchange couplings over multiple coordination shells.

%\begin{figure}[t]
%    \centering
%    \includegraphics[width=0.45\textwidth]{example-figure.pdf}
%    \caption{
%        Example figure caption. The caption should describe what is plotted,
%        the relevant parameters, and the physical message of the figure.
%    }
%    \label{fig:example}
%\end{figure}

%As shown in Fig.~\ref{fig:example}, the result illustrates the central behavior.

\section{Closing the loop between theory and experiment}
\label{sec:closing_loop}

The results presented above demonstrate that deep neural networks can
accurately reconstruct the microscopic Hamiltonian of a quantum magnet from
dispersion and intensities within a prescribed family of models, and that the
training process simultaneously diagnoses whether the supplied observables
contain enough information to constrain the parameters. The practical value of
this diagnostic, however, will also depend on how faithfully the synthetic
training data represent a real measurement. The invertibility analysis of
Sec.~\ref{sec:results} was carried out on idealized inputs: exact, delta-function magnon
branches $\omega_\pm(\mathbf{q})$ and bare spectral weights $I_\pm(\mathbf{q})$,
sampled on a uniform Brillouin-zone grid and corrupted only by additive Gaussian
noise. Real inelastic neutron-scattering (INS) data differ from this
idealization in ways that both distort and reduce the information available to
the network. 

In this section, we outline the main information-obscuring processes encountered in experiment. We further demonstrate that our approach remains effective even when two of the most important of these effects -- limited energy resolution and finite counting statistics -- are applied to the spectra, provided that the same effects are also applied to the training data. This result suggest that by incorporating experimental effects in the forward solver, a neural network can learn the inverse of the complete theory-to-experiment pipeline, providing a direct route from realistic neutron-scattering measurements to the underlying microscopic Hamiltonian.

\subsection{From the dynamical structure factor to measured intensities}
\label{sec:cl_intensities}

Rather than exact energies and intensities, $\omega_\pm(\mathbf{q})$ and
$I_\pm(\mathbf{q})$, a neutron scattering experiment instead measures a
continuous, four-dimensional field of intensities, $I(\bf q, \omega)$, that is proportional to the
neutron scattering cross section:
\begin{equation}
\frac{d^2\sigma}{d\Omega\,dE_f}
\;\propto\;
\frac{k_f}{k_i}\,\lvert f(\mathbf{q})\rvert^{2}\,e^{-2W}
\sum_{\alpha\beta}\!\left(\delta_{\alpha\beta}-\hat{q}_\alpha\hat{q}_\beta\right)
S^{\alpha\beta}(\mathbf{q},\omega).
\label{eq:cross_section}
\end{equation}
The structure factor itself, $S^{\alpha\beta}(\mathbf{q},\omega)$, here appears modified 
by several $\mathbf{q}$- and $\omega$-dependent factors: the kinematic factor
$k_f/k_i$; the magnetic form factor $\lvert f(\mathbf{q})\rvert^{2}$, which
suppresses signal at large $\lvert\mathbf{q}\rvert$; the Debye--Waller factor
$e^{-2W}$; and the polarization factor
$(\delta_{\alpha\beta}-\hat{q}_\alpha\hat{q}_\beta)$, which retains only the
spin-fluctuation components transverse to the momentum transfer~[51, 52]. For
the field-polarized state considered here, this last factor makes the observed
weight of a given magnon branch depend on the orientation of $\mathbf{q}$
relative to the applied field. At finite temperature, the detailed-balance
factor further reweights energy-gain and energy-loss scattering. 
Each of these factors is a known function of $\mathbf{q}$ and $\omega$ and can be applied deterministically to the forward-solver output. 
Although they can suppress or reweight parts of the dynamical response, thereby reducing the experimental sensitivity to some features of $\omega_{\pm}$ and $I_{\pm}$, they are all easily calculable and can thus be included in the generation of synthetic data for the purposes of performing an invertibility analysis, as described in Sec.~\ref{subsec:invertibility}. 
Their impact on the information available for Hamiltonian reconstruction is nevertheless expected to be less severe than that of incomplete momentum-energy coverage, finite instrumental resolution, and counting statistics, discussed in the following subsections.

\subsection{Data reduction, coverage, and resolution}
\label{sec:cl_reduction}

Modern single-crystal INS on direct-geometry time-of-flight spectrometers
records individual detection events, each tagged by detector position and time
of flight and, together with the sample orientation, mapped to a point in the
four-dimensional $(\mathbf{q},\omega)$ space. A complete dataset is assembled by
rotating the crystal through a sequence of orientations, and the region that can
be reached is bounded by the incident energy and by kinematic constraints, so
coverage of the Brillouin zone is generally neither uniform nor complete. The
accumulated events are then histogrammed onto a regular grid to produce an
estimate of $S(\mathbf{q},\omega)$~\cite{ewings2016horace, savici2022efficient}.
This binning is itself lossy: finite bins average the response
over regions of $(\mathbf{q},\omega)$ space, while the statistical quality of
the resulting bins can vary strongly across the measured volume.

Furthermore, the histogrammed intensity is a convolution of the true response with
the instrument resolution function, a $\mathbf{q}$- and $\omega$-dependent
(typically ellipsoidal) kernel set by the moderator pulse, choppers, and
detector geometry \cite{Violini2014, Granroth2015}. Extracting the sharp
dispersions $\omega_\pm(\mathbf{q})$ and integrated weights
$I_\pm(\mathbf{q})$ that the present networks take as input therefore requires
a further fitting step: cutting through the four-dimensional volume and fitting
resolution-convolved lineshapes (for example, damped-harmonic-oscillator
profiles) to locate peak centers and areas.
Finite coverage, binning, and resolution convolution reduce the
experimentally accessible information, while the subsequent extraction of
peak positions and spectral weights introduces additional, generally correlated,
uncertainties. These effects are not captured by adding independent Gaussian
noise to a clean spectrum.

\subsection{Modeling noise characteristics of an INS experiment}
\label{sec:cl_noise}

The dominant intrinsic uncertainty in an INS measurement is counting
statistics. Detected neutrons obey Poisson statistics, so the variance of a bin
equals its mean and the relative uncertainty scales as $N^{-1/2}$. This noise is
therefore heteroscedastic: weaker features, such as the low-intensity branch or
the high-$\lvert\mathbf{q}\rvert$ regions suppressed by the form factor, are
measured with proportionally larger error---in contrast to the homoscedastic
Gaussian model used in Secs.~III and IV, whose amplitude is tied to the global
bandwidth and maximum intensity of each spectrum. Superimposed on this random
component is a structured, correlated background arising from incoherent and
multiple scattering, phonons, the sample environment (for instance, Bragg peaks
from aluminum sample cans), and spurious features. Because such a background
contains systematic contributions that are not described by the
simple Poisson model for the magnetic signal, 
it is sensible to remove it as a separate reduction
step---using measured or modeled templates---rather than leave it for a network
to disentangle. Ideally, even imperfect subtraction would then leave residual
counting noise rather than a net bias. Alternatively, if a good model of
the correlated background is available,
it can be incorporated directly into the synthetic training data.
As these correlated background features generally depend on
very specific experimental conditions---including the sample,
sample environment, instrument, and instrument configuration---we disregard
them in the present study. In a material-specific application, however, such known
background contributions could be incorporated into the forward model in the
same way as the instrumental and statistical effects considered below.

% ---------------------------------------------------------------------
\begin{figure*}[t]
\includegraphics[width=\textwidth]{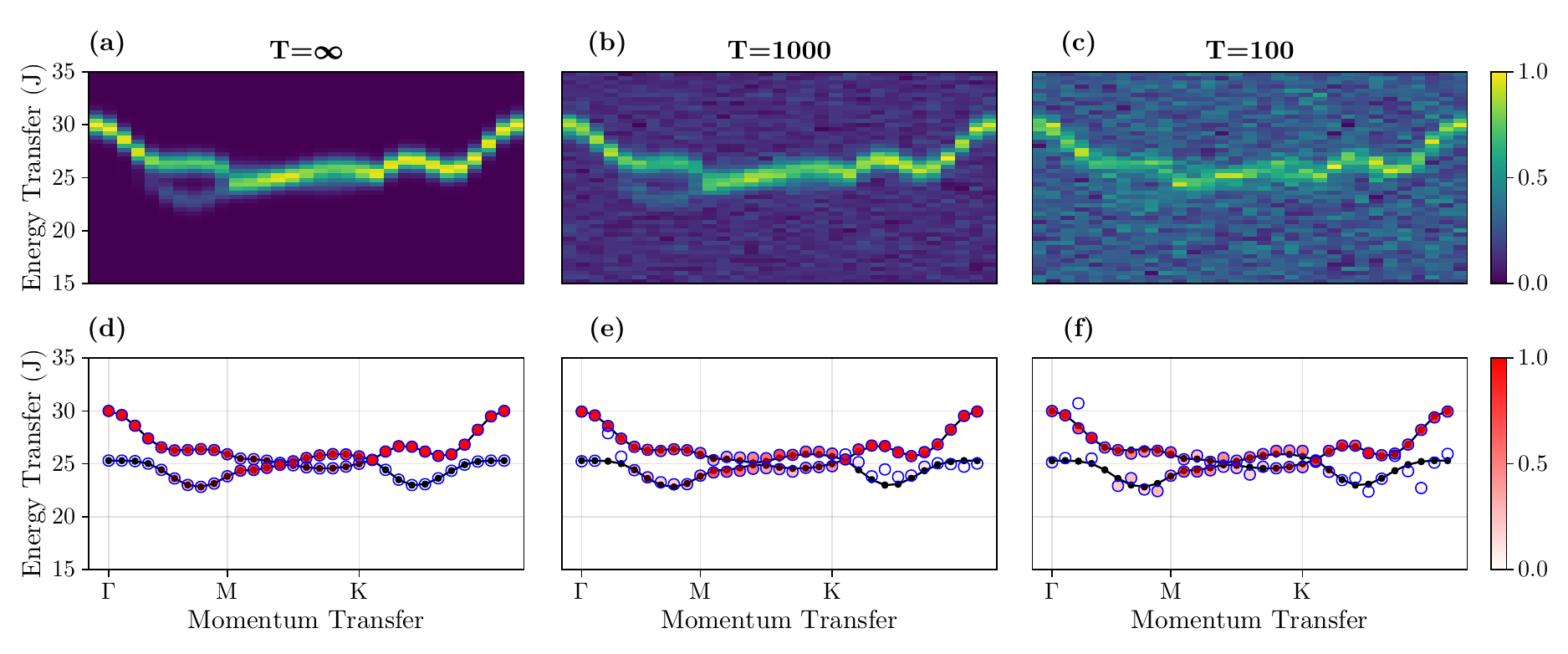}
\caption{\label{fig:poisson_maps}%
Effect of the counting-noise model at three exposures $T$. (a)--(c):
energy-convolved and binned intensity $S(\mathbf{q},\omega)$ along
the $\Gamma$--M--K path for $T=\infty$ (noise-free), $T=1000$, and
$T=100$. (d)--(f): the dispersions $\omega'_\pm(\mathbf{q})$ recovered
by the moment estimator, with marker size and color encoding the
recovered intensity $I'_\pm(\mathbf{q})$; black points mark the exact
branches. Counting noise blurs the intensity maps and scatters the
recovered branches, most strongly where the intensity---and hence the
local count rate---is smallest. The parameters used in this example are
$J_1=1$, $J_2=0.3407$, $J_3=0.3978$, $J_4=-0.5520$, $J_5=0.3407$,
$J_6=0.1658$, $J_7=0.4947$, and $J_8=0.2790$.}
\end{figure*}

\subsection{Network performance on noise model}
\label{sec:cl_sensitivity}

To demonstrate that the approach we have outlined remains viable when considering
the experimental factors of instrument resolution, histogramming, and counting
statistics, we replace the Gaussian noise injection of Sec.~\ref{sec:cl_noise}
by a simplified forward model of the measurement process itself.
For each Hamiltonian we (i)~first evaluate the exact
branches $\omega_\pm(\mathbf{q})$ and weights $I_\pm(\mathbf{q})$. (ii)~We then broaden
each branch into a Gaussian instrumental lineshape of fixed energy resolution
[$\mathrm{FWHM}=1.5\vert J_1\vert$, $\sigma=\mathrm{FWHM}/(2\sqrt{2\ln 2})$] on a uniform
$126$-point energy grid spanning $[0,50]$ in units of $\left\vert J_1\right\vert$. This yields an
intensity density at each $\mathbf{q}$ of
\begin{equation}
s\left(E\right) =
\sum_{n=\pm}\frac{I_n\left(\bf q\right)}{\sigma\sqrt{2\pi}}
\exp{\left(-\left(E-\omega_n\left(\bf q\right)\right)^2/2\sigma^2\right)}
\end{equation}
which is evaluated at the 126 bin-centers $E_k$.
(iii) We then treat the resulting binned
intensity density $s_k\equiv s(E_k)$ as the rate of an
inhomogeneous Poisson variable. Specifically, for each energy bin
$(\mathbf{q},E_k)$ we determine the number of neutron events as the sum of two Poisson distributed random variables, one with a rate determined by the intensity density, $n_k\sim\mathrm{Poisson}\left( s_k T\right)$, and the other with a uniform background rate of $g$, $m_k\sim\mathrm{Poisson}\left(gT\right)$; in both cases $T$ is understood as a unitless exposure duration, and $g=1$ throughout. The sampled intensity in each bin is finally given by
\begin{equation}
\tilde{s}_k=\left(n_k + m_k\right)/T - g.
\end{equation}
Note that $\tilde{s}_k \to s_k$ as $T\to\infty$.
In particular, the subtraction by $g$ ensures that the expected value of $\tilde{s}_k$ is 0 when $s_k$ is zero, but an individual sample may yield a negative value. This mimics the residual counting noise left by background subtraction. 
(iv)~Finally we recover perturbed estimates
$\omega'_\pm(\mathbf{q})$ and $I'_\pm(\mathbf{q})$ as follows. We consider two $\pm3\sigma$ windows about $\omega_+$ and $\omega_-$; where these bins overlap, we assign each to the nearer frequency ($\omega_+$ or $\omega_-$) to generate two disjoint regions.  Using these integration regions, we estimate the zeroth and first moments, $M_0=\sum \tilde{s}_k\Delta E$ and $M_1=\sum E_k \tilde{s}_k\Delta E$, and take $I'_\pm=M_0$ and $\omega'_\pm=M_1/M_0$. Where there is no intensity, we set $I'_\pm(\mathbf{q})=0$ and sample $\omega'_\pm(\mathbf{q})$ from a Gaussian distribution centered on $\omega_\pm(\mathbf{q})$ with $\sigma$ equal to the instrument resolution defined above. This is a simpler procedure than would be used in practice -- a full nonlinear fit to, for example, a mixture of Gaussians -- but captures the basic effects that counting noise introduces in the estimation of dispersion and intensity.

The single variable governing this model is
the exposure $T$, which plays the role of a measurement time: the expected number
of counts in a cell equals $T$ times the resolution-broadened intensity, so
$T\to\infty$ recovers the noise-free (resolution- and binning-limited) baseline,
while finite $T$ injects counting noise whose relative magnitude scales as
$T^{-1/2}$. For our study, we selected $T=100$ and $T=1000$
to represent two different counting-statistics regimes, with the
latter corresponding to a tenfold larger exposure.
Because both the recovered energies and weights are extracted from
the same counts, the perturbations to $\omega'_\pm$ and $I'_\pm$ are
induced jointly and self-consistently, unlike the independent Gaussian
perturbations of Sec.~III.
Consequently, the uncertainties in the extracted energies and
spectral weights are correlated and intensity dependent.
Fig.~\ref{fig:poisson_maps}  shows the effect: as $T$
decreases from $\infty$ to $1000$ and $100$, the binned $S(\mathbf{q},\omega)$
maps accumulate visible shot noise and the recovered dispersions scatter about
the true branches, most strongly on the weak branch and near the zone boundary
where the intensity---and hence the local count rate---is smallest.

As discussed in Sec.~\ref{subsec:accuracy_robustness}, the addition of Gaussian
noise to the training data functions to regularize the neural network and render
it more robust to noisy input data \cite{bishop1995training}. We therefore
consider two cases: first, we take the CNN1D model trained on $\sigma=0.05$
Gaussian noise and test its efficacy on data generated using the counting noise
model described above; additionally, we retrain
the CNN1D model on data generated with the counting process and test it on the Poisson data. When training and evaluating the model with Poisson noise, the same noise level T was used for both training and evaluation. Specifically, a model trained with $T = 1000$ was evaluated on data generated with $T = 1000$, while a model trained with $T = 100$ was evaluated on data generated with T = 100. For T = $\infty$, no Poisson noise was introduced during either training or evaluation. This noise-free case is equivalent to the Gaussian-noise setting with σ = 0 for both training and evaluation.

Table~\ref{tab:poisson} shows the results. Although the two schemes are
comparably accurate on noise-free input, the Gaussian-trained network degrades severely once genuine
counting noise is present, with the average relative error on the Random set
rising to $16.9\%$ at $T=1000$ and $17.2\%$ at $T=100$. Retraining on the
matching Poisson model restores the accuracy: the Random-set error falls to
$2.2\%$ and $3.9\%$ at the same two exposures---comparable to the clean-data
performance of Sec.~IV---while the sign of $J_1$ is classified with $100\%$
accuracy throughout.
The large improvement from $16.9\%$ to $2.2\%$ at $T=1000$
demonstrates that robustness requires not simply adding noise during training,
but reproducing the statistical structure of the measurement. At the parameter-resolved level, all seven continuous
couplings remain below $\sim\!3\%$ at $T=1000$; at the shorter $T=100$ exposure
only $J_3$ and $J_8$ exceed $5\%$ (at $5.8\%$ and $5.3\%$, respectively).

This simple example makes clear that it is essential to take into account the
statistical character of the noise in order to achieve successful
generalization. In the case of neutron measurements, this requires consideration
of resolution, binning and counting statistics.
Importantly, realistic counting noise does not appear to
constitute a fundamental obstacle to Hamiltonian reconstruction: when the
relevant measurement statistics are incorporated into the synthetic training
data, a straightforward neural network architecture retains the
representational capacity to perform the inverse problem.
We expect that the more completely the effects of
Secs.~\ref{sec:cl_intensities}--\ref{sec:cl_noise} are built into the training
database, the more directly the invertibility diagnostics of Sec.~IV can be read
as statements about a real measurement: for a given coverage, counting time,
and background level, the trained network can reveal whether the data contain
enough information to constrain the Hamiltonian and which additional
measurements would most improve the reconstruction. In this form the framework
becomes a quantitative tool for experiment design---for allocating beam time,
choosing incident energies and sample orientations, and identifying the regions
of reciprocal space that most sharply constrain the couplings---before the
measurement is performed.

% ---------------------------------------------------------------------
\begin{table}[t]
\caption{\label{tab:poisson}%
Average relative error (\%) of the CNN1D on Poisson-corrupted evaluation data,
for a network trained with the additive Gaussian scheme ($\sigma=0.05$) and one
trained directly on the Poisson counting-noise model, at three exposures $T$
($T=\infty$ is the noise-free reference, i.e. exact $\omega_\pm$ and
$I_\pm$). Errors are reported on the structured Test set and the independently
sampled Random set.}
\begin{ruledtabular}
\begin{tabular}{llccc}
Training noise & $T$ & $\mathrm{RE}_\mathrm{Test}$ & $\mathrm{RE}_\mathrm{Random}$ \\
\hline
Gaussian ($\sigma=0.05$) & $\infty$ & \phantom{0}1.8 & \phantom{0}2.6 \\
                         & $1000$   & 12.4           & 16.9           \\
                         & $100$    & 12.5           & 17.2           \\
Poisson (this work)      & $\infty$ & \phantom{0}0.6 & \phantom{0}0.9 \\
                         & $1000$   & \phantom{0}1.7 & \phantom{0}2.2 \\
                         & $100$    & \phantom{0}3.0 & \phantom{0}3.9 \\
\end{tabular}
\end{ruledtabular}
\end{table}

% ---------------------------------------------------------------------
\section{Discussion}
\label{sec:discussion}

The results of this work establish that Hamiltonian reconstruction from
neutron-scattering data can be formulated as a supervised-learning problem
within a prescribed family of microscopic models. Once the relevant model
class has been identified, the forward solver generates the dynamical
responses throughout its parameter space, and the neural network learns
the corresponding inverse relation. The successful reconstruction of
independently sampled Hamiltonians demonstrates that the network is not
simply identifying members of the training set, but interpolating the
inverse map between them. The fact that three substantially different
network architectures achieve comparable accuracy further indicates that
this conclusion is not tied to a particular neural-network representation.

A central lesson of the present study is that the success of Hamiltonian
reconstruction is controlled not only by the learning algorithm, but also
by the information content of the measured response. The comparison
between dispersion-only and dispersion-plus-intensity reconstructions
provides a direct example. Distinct Hamiltonians can have nearly
indistinguishable magnon dispersions and therefore cannot be reliably
distinguished from excitation energies alone. The spectral weights contain
additional eigenvector information and can remove these ambiguities. More
generally, systematic failures of the learned inverse map can reveal that
the available observables do not sufficiently constrain the microscopic
model. Supervised learning can therefore serve simultaneously as an
inference method and as a diagnostic of whether a proposed measurement
contains the information required to solve the inverse problem.

The stability of the reconstruction against noise provides a second
important indication of the practical viability of this approach. The
prediction error increases continuously as uncertainty is added to the
spectral data, rather than exhibiting an abrupt breakdown of the inverse
mapping. Moreover, incorporating comparable uncertainty into the synthetic
training data substantially improves the reconstruction of noisy inputs.
This observation suggests a natural strategy for applications to
experiment: the statistical and instrumental uncertainties expected in a
particular measurement can be incorporated into the generation of the
training database, allowing the learned inverse model to be adapted to the
quality of the experimental data.

The computational structure of the approach is also fundamentally
different from conventional Hamiltonian fitting. In an iterative fit, the
forward problem must be solved repeatedly for each material and for each
trial set of parameters. Here, the computationally demanding exploration
of parameter space is performed once, during construction of the synthetic
database. The resulting trained network can subsequently infer Hamiltonian
parameters at negligible computational cost compared with the forward
calculation. In this sense, the trained inverse model provides a portable
representation of the information generated by the forward solver. A
training database produced using a sophisticated theoretical method can
therefore be reused for many materials belonging to the same model family,
without requiring every user to implement or repeatedly execute the
underlying solver.

This perspective is particularly relevant for the dissemination of
theoretical capabilities within the neutron-scattering community. Accurate
forward calculations may require specialized numerical methods, substantial
computational resources, and expertise that are not uniformly available
to experimental groups. Once the corresponding inverse model has been
trained, however, its deployment requires only the measured spectral
information. Machine learning therefore provides a possible mechanism for
turning specialized theoretical calculations into broadly accessible
analysis tools. The role of the neural network is not to replace the
forward solver or the physical modeling used to define the Hamiltonian
family, but to make the information generated by those calculations
reusable and readily accessible.

The present proof of principle also makes clear the limitations of this
strategy. The neural network can only infer parameters within the model
family represented in its training database, and its predictions can only
be as reliable as the forward solver used to generate that database. A
successful reconstruction therefore does not by itself establish that the
assumed Hamiltonian family provides an adequate description of a real
material. Conversely, experimental spectra containing interactions or
physical effects absent from the training set may fall outside the domain
in which the learned inverse relation is reliable. Identifying such
out-of-distribution data and quantifying the corresponding uncertainty will
be essential for applications to experimental systems.

Although we have chosen the fully polarized Heisenberg model because its
zero-temperature transverse dynamical spin structure factor provides an
exact and computationally inexpensive benchmark, the learning framework
itself does not depend on linear spin-wave theory. Any theoretical method
capable of generating sufficiently accurate dynamical response functions
over the relevant Hamiltonian parameter space can, in principle, be used
to construct the training database. The principal challenge in extending
the approach to more strongly correlated quantum magnets is therefore not
conceptual, but computational: sufficiently large and representative
training sets must be generated using more demanding forward solvers.
Progress in numerical many-body methods and computational resources should
progressively enlarge the range of magnetic materials for which this
strategy becomes practical.

Taken together, these observations suggest that machine learning can play
a broader role in inverse neutron scattering than simply accelerating
parameter fitting. It provides a framework in which microscopic modeling,
forward calculations, information-content analysis, and Hamiltonian
reconstruction are integrated into a common workflow. The present results
demonstrate this principle in a controlled setting and provide a foundation
for extending it to increasingly realistic neutron-scattering problems.

\section{Conclusion}
\label{sec:conclusion}

We have demonstrated that the inverse neutron-scattering problem can be
formulated and solved as a supervised-learning problem within a prescribed
family of microscopic Hamiltonians. Using an eight-parameter family of
honeycomb Heisenberg models for which the zero-temperature transverse
dynamical spin structure factor can be computed exactly, we constructed a
large synthetic training database and showed that neural networks can
accurately reconstruct Hamiltonians that were not included in the training
set. The reconstruction remains stable in the presence of noise, and
comparable performance across three substantially different neural-network
architectures indicates that the result is not tied to a particular
network representation.

Beyond Hamiltonian reconstruction itself, our results demonstrate that the
learning process can be used to assess whether the available
neutron-scattering information sufficiently constrains the inverse problem.
In particular, we found that magnon dispersions alone can leave distinct
Hamiltonians nearly indistinguishable, whereas the corresponding spectral
weights provide additional eigenvector information that resolves these
ambiguities. The same strategy can therefore be used more generally to
identify which experimentally accessible observables are required for
reliable Hamiltonian reconstruction, providing a direct connection between
inverse modeling and the design of neutron-scattering experiments.

We further showed that the inverse reconstruction remains
accurate in the presence of realistic counting statistics when the
corresponding measurement process is incorporated into the synthetic training
data, providing a direct route toward applications to experimental INS data.

The framework also changes the computational organization of Hamiltonian
determination. The expensive exploration of parameter space is performed
once, during construction of the synthetic training database, while the
trained inverse model subsequently provides essentially instantaneous
Hamiltonian estimates. This makes it possible to reuse the information
generated by sophisticated forward solvers and provides a route for making
specialized theoretical capabilities broadly accessible to the
neutron-scattering community.

 \section*{Code Availability}

The code for data generation, model training and evaluation,
and figure generation is publicly available on GitHub at
\url{https://github.com/Jingyi-Luo/inverse-neutron-ML}
and permanently archived on Zenodo
\cite{luo_2026_22752569}.

\begin{acknowledgments}
We thank Daniel Pajerowski for useful discussions. Work at the University of Tennessee was supported by the National Science Foundation Materials Research Science and Engineering Center program through the UT Knoxville Center for Advanced Materials and Manufacturing (DMR-2309083). 
D.D. was supported by the Scientific User Facilities Division, Office of Basic Energy Sciences, U.S. Department of Energy, under contract no. DE-AC0500OR22725 with UT Battelle, LLC.
% DD was supported by  U.S. Department of Energy, Office of Science, Basic Energy Sciences (BES) under the Genesis Mission BES AI Pathfinder Program, MAIQMag: Multimodal AI for 2D Quantum Magnets. 
\end{acknowledgments}

% ---------- Appendices ----------
\appendix

\section{One-magnon spectrum and transverse structure factor above the saturation field}
\label{app:Derivation}

\subsection{One-magnon energies}
\label{section:energies}

Let $|\Psi_0\rangle$ be the fully polarized state. 
The Holstein--Primakoff representation is
\begin{align}
    S_{\bf R}^{+,\alpha}
    &=\sqrt{2S}
    \left(1-\frac{n_{{\bf R},\alpha}}{2S}\right)^{1/2}
    b_{{\bf R},\alpha},\\
    S_{\bf R}^{-,\alpha}
    &=\sqrt{2S}\,b_{{\bf R},\alpha}^{\dagger}
    \left(1-\frac{n_{{\bf R},\alpha}}{2S}\right)^{1/2},\\
    S_{\bf R}^{z,\alpha}
    &=S-n_{{\bf R},\alpha},
    \qquad
    n_{{\bf R},\alpha}
    =b_{{\bf R},\alpha}^{\dagger}b_{{\bf R},\alpha}.
\end{align}
Within the zero- and one-magnon sectors,
$S_{\bf R}^{+,\alpha}$ and $S_{\bf R}^{-,\alpha}$ act as
$\sqrt{2S}\,b_{{\bf R},\alpha}$ and
$\sqrt{2S}\,b_{{\bf R},\alpha}^{\dagger}$, respectively. 
Using these relations in Eq.~\eqref{eq:H} and omitting the ground-state energy gives
\begin{align}
    H_{\rm sw}
    ={}&-S\sum_{{\bf R},\alpha}
    \left(\sum_{{\bf r},\beta}J_{\bf r}^{\alpha\beta}\right)
    b_{{\bf R},\alpha}^{\dagger}b_{{\bf R},\alpha}
    \nonumber\\
    &+\frac S2\sum_{{\bf R},{\bf r}}
    \sum_{\alpha,\beta}J_{\bf r}^{\alpha\beta}
    \left(
    b_{{\bf R},\alpha}^{\dagger}b_{{\bf R}+{\bf r},\beta}
    +b_{{\bf R}+{\bf r},\beta}^{\dagger}b_{{\bf R},\alpha}
    \right)
    \nonumber\\
    &+h\sum_{{\bf R},\alpha}
    b_{{\bf R},\alpha}^{\dagger}b_{{\bf R},\alpha}.
    \label{eq:honey_sw_realspace}
\end{align}
The first term is an on-site contribution proportional to the exchange row sum for each sublattice. 
We use the Fourier convention
\begin{equation}
    b_{{\bf R},\alpha}
    =\frac1{\sqrt{N_c}}\sum_{\bf q}
    e^{2\pi i{\bf q}\cdot{\bf R}}b_{{\bf q},\alpha},
\end{equation}
where $N_c$ is the number of primitive cells and momenta are expressed in reciprocal-lattice units.

To define the exchange matrix in the convention used by the forward solver, let $\mathcal R_{AA}^{+}$ contain one displacement from each inversion-related pair of same-sublattice bonds, and let
$\mathcal R_{AB}$ contain the $A$-to-$B$ bonds. 
Define
\begin{equation}
    \begin{aligned}
    J^{AA}({\bf q})
    &=\sum_{{\bf r}\in\mathcal R_{AA}^{+}}
    J_{\bf r}^{AA}e^{2\pi i{\bf q}\cdot{\bf r}},\\
    J^{AB}({\bf q})
    &=\sum_{{\bf r}\in\mathcal R_{AB}}
    J_{\bf r}^{AB}e^{2\pi i{\bf q}\cdot{\bf r}}.
    \end{aligned}
    \label{eq:Jq}
\end{equation}
The full same-sublattice Fourier sum is therefore $2\operatorname{Re}J^{AA}({\bf q})$. 
Because $A$ and $B$ are equivalent sublattices in this model,
\begin{equation}
    J({\bf q})=
    \begin{pmatrix}
    2\operatorname{Re}J^{AA}({\bf q}) & J^{AB}({\bf q})\\
    [J^{AB}({\bf q})]^* & 2\operatorname{Re}J^{AA}({\bf q})
    \end{pmatrix},
\end{equation}
and
\begin{equation}
    D=\left[2J^{AA}({\bf 0})+J^{AB}({\bf 0})\right]I_2.
    \label{eq:Dmatrix}
\end{equation}

Fourier transformation of Eq.~\eqref{eq:honey_sw_realspace} now gives Eq.~\eqref{eq:Mq}. 
Its normalized eigenvectors satisfy
\begin{equation}
    [J({\bf q})-D]{\bf u}_{\bf q}^{(n)}
    =\mu_n({\bf q}){\bf u}_{\bf q}^{(n)},
    \qquad
    \omega_n({\bf q})=h+S\mu_n({\bf q}).
    \label{eq:eigenproblem}
\end{equation}
Writing
$J^{AB}({\bf q})=|J^{AB}({\bf q})|e^{i\phi_{\bf q}}$, a convenient choice of eigenvectors away from degeneracies is
\begin{equation}
    {\bf u}_{\bf q}^{(\pm)}
    =\frac1{\sqrt2}
    \begin{pmatrix}
    1\\
    \pm e^{-i\phi_{\bf q}}
    \end{pmatrix}.
\end{equation}
These give the two dispersions stated in Sec.~\ref{subsec:forward}.

\subsection{Transverse dynamical spin structure factor}
\label{section:DSSF}

We define the momentum-space spin-lowering operator with the physical positions of the two basis sites:
\begin{equation}
    S^-({\bf q})
    =\frac1{\sqrt{2N_c}}
    \sum_{{\bf R},\alpha}
    e^{2\pi i{\bf q}\cdot({\bf R}+{\bf d}_\alpha)}
    S_{\bf R}^{-,\alpha},
    \,
    S^+({\bf q})=[S^-({\bf q})]^\dagger.
\end{equation}
The factor $1/\sqrt{2N_c}$ normalizes the structure factor per spin. In the one-magnon sector,
\begin{equation}
    S^-({\bf q})|\Psi_0\rangle
    =\sqrt S\sum_\alpha
    e^{2\pi i{\bf q}\cdot{\bf d}_\alpha}
    b_{{\bf q},\alpha}^{\dagger}|\Psi_0\rangle.
\end{equation}

Define the magnon creation operators and states by
\begin{equation}
    \beta_{{\bf q},n}^{\dagger}
    =\sum_\alpha u_{{\bf q},\alpha}^{(n)}
    b_{{\bf q},\alpha}^{\dagger},
    \qquad
    |n,{\bf q}\rangle
    =\beta_{{\bf q},n}^{\dagger}|\Psi_0\rangle.
\end{equation}
Unitarity of the eigenvector matrix gives
$b_{{\bf q},\alpha}^{\dagger}
=\sum_n u_{{\bf q},\alpha}^{(n)*}
\beta_{{\bf q},n}^{\dagger}$.
The matrix element is therefore
\begin{equation}
    \langle n,{\bf q}|S^-({\bf q})|\Psi_0\rangle
    =\sqrt S\sum_\alpha
    e^{2\pi i{\bf q}\cdot{\bf d}_\alpha}
    u_{{\bf q},\alpha}^{(n)*}.
\end{equation}

For the zero-temperature transverse structure factor,
\begin{equation}
    S^{+-}({\bf q},\omega)
    =\frac1{2\pi}\int_{-\infty}^{\infty}dt\,
    e^{i\omega t}
    \langle\Psi_0|
    S^+({\bf q},t)S^-({\bf q},0)
    |\Psi_0\rangle,
\end{equation}
the spectral weight of branch $n$ is
\begin{equation}
    I_n({\bf q})
    =S\left|
    \sum_\alpha
    e^{2\pi i{\bf q}\cdot{\bf d}_\alpha}
    u_{{\bf q},\alpha}^{(n)*}
    \right|^2.
\end{equation}
Substituting the two eigenvectors and setting
${\bf d}={\bf d}_B-{\bf d}_A$ yields
\begin{equation}
    I_\pm({\bf q})
    =S\left[
    1\pm\cos\!\left(
    2\pi{\bf q}\cdot{\bf d}+\phi_{\bf q}
    \right)\right],
\end{equation}
as used in Sec.~\ref{subsec:forward}. 
The weights sum to $I_+({\bf q})+I_-({\bf q})=2S$, the total transverse weight per spin.

% ---------- Bibliography ----------
\clearpage
\bibliography{ref_honeycomb}

\end{document}